\documentclass[scriptaddress,pre,onecolumn,11pt]{revtex4-1}
\usepackage{mathrsfs}
\usepackage{amsmath,amssymb}
\usepackage{bm}
\usepackage[dvipdfmx]{graphicx,color}
\usepackage{url}

\begin{document}

\title{Opinion formation with upper and lower bounds}
\author{Ryosuke Yano}
\affiliation{Department of Advanced Energy, University of Tokyo, 5-1-5 Kashiwanoha, Kashiwa, Chiba 277-8561, Japan}
\email{yano@k.u-tokyo.ac.jp}
\author{Arnaud Martin}
\affiliation{Sino-French Institute for Nuclear Engineering and Technology, Sun Yat-sen University, Tangjiawan, Xiangzhou, City of Zhuhai 519082, Guangdong province, People's Republic of China}
\email{amphyschim@gmail.com}
\begin{abstract}
We investigate the opinion formation with upper and lower bounds. We formulate the binary exchange of opinions between two peoples under the second (or political) party using the relativistic inelastic-Boltzmann-Vlasov equation with randomly perturbed motion. In this paper, we discuss the relativistic effects on the opinion formation of peoples from the standpoint of the relativistic kinetic theory.
\end{abstract}
\maketitle
\section{Introduction}
The opinion formation has been studied with great interests in the framework of sociophysics \cite{Galam} \cite{Naldi}. In previous studies, microscopic models have been proposed by Sznajd \cite{Sznajd} or Hegselman and Krause \cite{Krause}. In particular, the master equation, which describes the microscopic motions of opinions proposed by Hegselman and Krause, indicates some clustering states of opinions as a result of averaging of opinions, which are similar to each other. As an advanced model, the hierarchy of opinions was proposed to demonstrate the opinion formation in democratic parties by Galam \cite{Galam}. Meanwhile, the kinetic model of the opinion formation has been studied by Toscani and his coworkers \cite{Toscani} using the inelastic Boltzmann equation or partial differential equation (PDE), which corresponds to Sznajd \cite{Sznajd} model in Ochrombel simplification \cite{Ochrombel} on a complete graph \cite{Slanina}. The Fokker-Planck type equation by Toscani is written as \cite{Toscani}
\begin{eqnarray}
\frac{\partial f\left(\tau,m\right)}{\partial \tau}=\frac{\lambda}{2}\frac{\partial^2 \varphi(m) f\left(\tau,m\right)}{\partial m^2}+\frac{\partial \left(m-\bar{m}\right)f\left(\tau,m\right)}{\partial m},
\end{eqnarray}
where $\tau=\left(1+\Lambda\right)/2 t$ ($\Lambda$: restitution coefficient, $t \subseteq \mathbb{R}^+$: time), $m$ is the value of opinion defined by $m \in M ~\left\{x \in M| |x| < 1 \cap x \in \mathbb{R}\right\}$, $\bar{m}$ is the mean opinion, $f\left(\tau,m\right) \subseteq \mathbb{R}^+ \times M$ is the distribution function of the opinion, and $\lambda \subseteq \mathbb{R}^+$ is related to the the temperature of the opinion. Provided that $\varphi(m)=1$ in the right hand side of Eq. (1), Eq. (1) is equivalent to the classical Fokker-Planck equation \cite{Yano2}. The diffusion term, which is the first term in the right hand side of Eq. (1), corresponds to the heating term and the convection term, which is the second term in the right hand of Eq. (1), corresponds to the cooling term, which expresses the convergence of $m$ to $\bar{m}$.\\
Toscani calculated steady solutions of Eq. (1) for $\varphi(m)=\left(1-m^2\right)^2$, $\left(1-|m|\right)^2$ and $1-m^2$ as
\begin{eqnarray}
f_\infty(m)&=&A_1 \left(1+m\right)^{-2+\frac{m}{2\lambda}}\left(1-m\right)^{-2-\frac{m}{2\lambda}}\exp\left(-\frac{1-\bar{m}m}{\lambda\left(1-m^2\right)}\right),\nonumber~~~~\mbox{for}~~\varphi(m)=\left(1-m^2\right)^2,\\
&=& A_2 \left(1-|m|\right)^{-2-\frac{2}{\lambda}}\exp\left(-\frac{1-\bar{m}m/|m|}{2\lambda \left(1-|m|\right)}\right),~~~~\mbox{for}~~\varphi(m)=\left(1-|m|\right)^2,\nonumber \\
&=& A_3 \left(\frac{1}{1+m}\right)^{1-\frac{1+\bar{m}}{\lambda}} \left(\frac{1}{1-m}\right)^{1-\frac{1-\bar{m}}{\lambda}},~~~~\mbox{for}~~\varphi(m)=1-m^2,
\end{eqnarray}
where $A_i$ ($i=1,2,3$) are factors for the normalization.\\
Toscani set the restriction on the diffusion term using $\varphi(m)$, whereas we can consider other formulations of the Fokker-Planck type equation, which holds the causality of $m$. For example, the linear $q$-Fokker-Planck equation sets the upper and lower bounds to $m$ \cite{Yano2}, whereas the mathematical characteristics of the linear $q$-Fokker-Planck equation requires further discussions including the calculation of moments such as $q$-averaged moments or normal averaged moments \cite{Yano2}.\\
As another Fokker-Planck type equation, we can consider following model equation:
\begin{eqnarray}
\frac{\partial f\left(\tau,m\right)}{\partial \tau}={\lambda}\frac{\partial}{\partial m}\left(1-m^2\right)\frac{\partial f\left(\tau,m\right)}{\partial m}+\frac{\partial \left(1-m^2\right)^{-\frac{1}{2}}\left(1-\bar{m}^2\right)^{-\frac{1}{2}}\left(m-\bar{m}\right)f\left(\tau,m\right)}{\partial m},
\end{eqnarray}
Equation (3) also satisfies causality $|m|<1$. The steady solution of Eq. (3) is obtained as
\begin{eqnarray}
f_\infty(m)=A_4 \exp\left[-\frac{1}{\lambda}\left(1-m^2\right)^{-\frac{1}{2}}\left(1-\bar{m}^2\right)^{-\frac{1}{2}}\left(1-m\bar{m}\right)\right],
\end{eqnarray}
where $A_4$ is a factor for the normalization.\\
$f_\infty(m)$ in Eq. (4) is equivalent to one dimensional (1D) Maxwell-J$\ddot{\mbox{u}}$ttner function, which reveals the thermal equilibrium distribution of the relativistic gas, \textcolor{red}{when $A_4=n/(2K_1(\theta^{-1}))$ ($n$: number density of opinions, $K_n$: the $n$-th order modified Bessel function of the second kind, $\theta$: temperature in the opinion system) and $\lambda=\theta$}.\\
Equation (4) indicates that 1D Maxwell-J$\ddot{\mbox{u}}$ttner function might be one of possible solutions in the opinion formation with upper and lower bounds, when we choose parameters of the diffusion and convection rates in the Fokker-Planck type equation to hold the causality of $m$. Actually, the form of the power number of the exponential in $f_\infty(m)$, which is obtained using $\varphi(m)=\left(1-m^2\right)^2$ or $\left(1-|m|\right)^2$ in Eq. (2), is similar to that in 1D Maxwell-J$\ddot{\mbox{u}}$ttner function. Thus, we adopt the relativistic inelastic-Boltzmann equation to consider the opinion formation with upper and lower bounds. Of course, the application of the relativistic motion of the opinion seems to be too extensive, because the phase space $x:=\int m(t) dt \in \mathbb{R}$ is not postulated in the opinion formation. We, however, assume that the opinion $m$ is always uniformly distributed in the phase space $x$. Therefore, we can always neglect the phase space $x$. On the other hand, we must consider the Lorentz contraction in time $t$. The larger $|m|$ means the slower time evolution of $f(m)$ owing to the Lorentz contraction. Such an interpretation of the time evolution of $f(m)$ is reflected by $\varphi(m)$ in Eq. (1), as indicated by the fact that the increase of $|m|$ yields to the slower diffusion rate owing to the characteristics of $\varphi(m)$. The advantage of the use of the relativistic inelastic-Boltzmann equation over the conventional kinetic model in Eq. (1) or linear $q$-Fokker-Planck equation is that the thermodynamic characteristics of the relativistic gas such as H theorem, heat capacity, and so on, have been studied in previous studies, strictly \cite{Cercignani}.\\
The binary exchange of opinions between two peoples is expressed using the relativistic inelastic-Boltzmann equation, whereas the diffusion via the self-thinking is expressed by incorporating randomly perturbed motion \cite{Bassetti} into the binary inelastic collision, because the self-thinking usually occurs at the binary exchange of opinions between two peoples. The effects of the second (or political) party on the opinion formation of peoples can be expressed using the Vlasov term, when we regard the second (or political) party as an external field, which has the influence on an opinion of a people through an external force. In this case, opinion dynamics of peoples are expressed using the relativistic inelastic-Boltzmann-Vlasov equation. The analytical discussion of the relativistic inelastic-Boltzmann-Vlasov equation indicates that relativistic effects are markedly significant for the opinion formation of peoples, as discussed in Sec. II. In addition to Lorentz factor, the thermally relativistic measure $\chi=\mathcal{M}/(\theta)$ ( $\mathcal{M}$: mass of an opinion, $\theta$: temperature) is a significant parameter, which characterizes the cooling rate of $\theta$ via the inelastic collisions or external force by the second (or political) party. Here, the mass of opinion characterizes the strength of the opinion. Therefore, the mass of opinion of the leader in the (political) community must be larger than that of the individual, who is not interested in the (political) issue. Here, we assume that the mass of all the individuals is common, because we regard the second (or political) party as the external field. As with the temperature of the opinion system, we consider it in later discussions. Finally, numerical results indicate that the opinion formation strongly depends on the thermally nonequilibrium state, as discussed in Sec. III.\\
This paper is organized as follows. In Sec. II, we propose the relativistic inelastic-Boltzmann-Vlasov equation to express the opinion formation of peoples by assuming that dynamics of opinions follows those of inelastic hard spheres, and discuss the cooling rate of $\theta$ under the thermally equilibrium state. In Sec. III, we discuss numerical results, which are obtained by solving the relativistic inelastic-Boltzmann-Vlasov equation. Finally, we make concluding remarks in Sec. IV. 
\section{Relativistic inelastic-Boltzmann-Vlasov equation and its characteristics}
Before formulating the relativistic inelastic-Boltzmann-Vlasov equation, we must describe some postulations to relate the relativistic inelastic-Boltzmann-Vlasov equation with the dynamics of opinions. Firstly, the interaction of opinions among multiple peoples beyond two peoples is neglected, because we restrict ourselves to the binary exchange of opinions between two peoples. Consequently, one dimensional Boltzmann equation is considered as a kinetic model. On the other hand, one dimensional elastic Boltzmann equation never change the state of opinions, because the elastic collision of two opinions means the exchange of two opinions, namely, $m \rightarrow m_\ast$ and $m_\ast \rightarrow m$, when the mass of opinion is common for all the peoples. Therefore, we consider one dimensional inelastic-Boltzmann equation to express the binary exchange of two opinions. As a result of inelastic collisions, two opinions move toward compromised opinions of two opinions. Additionally, we assume that the potential between two opinions follows hard sphere potential. Finally, an opinion can be regarded as a inelastic hard sphere (IHS) \cite{Santos} with mass $\mathcal{M}$ and diameter $d$. Here, $d$ expresses the range of the influence of the opinion. Thus, $d$ of the leader of the (political) community must be larger than that of the individual, who is not interested in the (political) issue. Meanwhile, we assume that $d$ of all the individuals are common, because we regard the second (political) party as the external field. The collision frequency of two opinions ($m$ and $m_\ast$) are proportional to the product of the relative magnitude of two opinions ($g=|m-m_\ast|$) with the collisional cross section of hard spheres, namely, $\sim g \pi d^2$. The collision frequency increases in accordance with the increase of $g$. Therefore, two opinions, which are more different, collide with each other more frequently. Meanwhile, microscopic model by Hegselman and Krause \cite{Krause} indicates that an opinion ($m$) interacts with an opinion ($m_\ast$), when $g \le \epsilon$ ($\epsilon$: constant). As a result, the description of the opinion formation by one dimensional inelastic-Boltzmann equation for the IHS is markedly different from that by the microscopic model by Hegselman and Krause \cite{Krause}. Meanwhile, the process of the opinion formation surely depends on the scale of the system, namely, from the national scale to a small party in the company. In our future study, we must investigate what kinetic model is the best to demonstrate the opinion formation in accordance with the scale of the system. Of course, the tendency of swarming (clustering) of peoples with similar opinions via the microscopic model by Hegselman and Krause \cite{Krause} might be plausible in a large system, when the system is hierarchic enough to neglect an opinion of one people, whose opinion is quite different from that of the majority. Meanwhile, we favor the more frequent exchanges of opinions between two individuals, whose opinion is markedly different, when all the opinions are exchanged without the hierarchy (i.e., small social community). Less frequency between two individuals, who have more different opinions, can be expressed by the soft potential such as Coulomb potential, where the collision frequency depends on $1/|m-m_\ast|^\delta$ ($0<\delta$).\\
The strength of opinions are bounded by $|m| \le 1$. As a result of a causality of $m$, the kinetic model, which always satisfies the causality of $m$ is the one dimensional relativistic kinetic model, where two peoples exchange momentums of two opinions, which are expressed with $p=\mathcal{M}\gamma(m)m$ and $p_\ast=\mathcal{M}\gamma(m_\ast)m_\ast$, \textcolor{red}{in which $\gamma(m):=1/\sqrt{1-m^2}$ (or $\gamma(m_\ast):=1/\sqrt{1-m_\ast^2}$) is the Lorentz factor}. Two peoples exchange "momentums" of two opinions. Consequently, the momentum of the strong opinion ($|m| \sim 1$) is emphasized by the Lorentz factor at the binary exchange of opinions between two peoples. In the relativistic motion of the IHS, the relative magnitude of two opinions are expressed using M\o ller's relative velocity \cite{Cercignani}.\\
In the nonrelativistic gas, the inelastic collision of two particles leads to the convergence of velocities of all the particles to initial mean velocity under the spatially homogeneous state, whereas such a convergence of all the opinions to initial mean opinion never be guaranteed in the relativistic state, as discussed in later Remark 1, whereas all the opinions converge to a specific opinion owing to the inelastic collision. To avoid such a convergence of all the opinions to the specific opinion, we consider randomly perturbed motion of two opinions at the binary exchange of momentums of opinions between two peoples. Hereafter, the mass of opinion is fixed to unity, namely, $\mathcal{M}=1$ for simplicity, \textcolor{red}{from which $\chi=1/\theta$ is obtained}.\\
Randomly perturbed terms are added to momentums at the binary exchange of momentums of opinions between two peoples to satisfy the conservation of total momentum of two colliding opinions. Here, randomly perturbed motion corresponds to the Brownian motion at the binary collision, which is constrained by the conservation of two momentums. Meanwhile, we numerically investigate, whether randomly perturbed motion at the binary collision always yields the heating or not, in later discussion.\\
Finally, the relativistic inelastic-Boltzmann-Vlasov equation is formulated as
\begin{eqnarray}
&&p^0\frac{\partial f\left(t,p\right)}{\partial x^0} \nonumber \\
&&=A \int_{-\infty}^{\infty} \left[\frac{1}{\mathcal{J}}f\left(t,p^{\prime\prime}\right) f\left(t,p_\ast^{\prime\prime}\right)-f\left(t,p\right) f\left(t,p_\ast\right)\right] F \frac{dp_\ast}{p^0_\ast} \nonumber \\
&&+B p^0 \frac{\partial \left(p-P\right)f\left(t,p\right)}{\partial p},
\end{eqnarray}
where $f\left(t,p\right) \subseteq \mathbb{R}^+ \times \mathbb{R}$ is the distribution function, $t$ is the time, $x^0=t$, and $p^0=1/\sqrt{1-m^2}$. The first term in the right hand side of Eq. (5) corresponds to relativistic inelastic-collisions with randomly perturbed motion, where $F=g_{\o}/\left(p^0p^0_\ast\right)$ ($g_{\o}$: M\o ller's relative velocity) \cite{Cercignani}, whereas the second term in the right hand side of Eq. (5) corresponds to the term, which expresses the external force on an opinion of a people by the second (or political) party, where $P=m_p\gamma\left(m_p\right)$ and $m_p$ is the strength of the opinion of the second ( or political) party. The rate of the binary exchange of opinions between two peoples and rate of the concentration of $m$ to $m_p$ via the second (or political) party are expressed by $A$ and $B$ in Eq. (5), respectively. As a result of the direct relativistic inelastic-collision with randomly perturbed motion, momentums of two colliding opinions, namely, $p$ and $p_\ast$, change to $p^\prime$ and $p_\ast^\prime$, which are defined by
\begin{eqnarray}
&&p^\prime=p+\frac{1+\Lambda}{2}\left(p_\ast-p+\Delta\left(p,p_\ast\right)\right)\\ 
&&p_\ast^\prime=p_\ast-\frac{1+\Lambda}{2}\left(p_\ast-p+\Delta\left(p,p_\ast\right)\right),
\end{eqnarray}
where $\Lambda$ is the inelasticity coefficient ($0 \le \Lambda \le 1$), and $\Delta$ is randomly perturbed motion via the self-thinking. On the other hand, momentums of two colliding opinions, namely, $p^{\prime\prime}$ and $p_\ast^{\prime\prime}$, change to $p$ and $p_\ast$, in which $p^{\prime\prime}$ and $p_\ast^{\prime\prime}$ are defined by
\begin{eqnarray}
&&p^{\prime\prime}=p+\frac{1+\Lambda}{2\Lambda}\left(p_\ast-p+\Delta\left(p,p_\ast\right)\right), \nonumber\\ 
&&p_\ast^{\prime\prime}=p_\ast-\frac{1+\Lambda}{2\Lambda}\left(p_\ast-p+\Delta\left(p,p_\ast\right)\right).
\end{eqnarray}
Consequently, the total momentum is conserved by the relativistic inelastic-collision with randomly perturbed motion, whereas the total energy ($E+E_\ast=\sqrt{1+p^2}+\sqrt{1+p_\ast^2}$) is not conserved by the relativistic inelastic-collision with randomly perturbed motion. Finally, $\mathcal{J}$ in Eq. (5) is the Jacobian, which is defined by
\begin{eqnarray}
J:= |\mbox{det} \partial (p^{\prime\prime},p_\ast^{\prime\prime})/\partial \left(p,p_\ast\right)|^{-1}=\left|1/\Lambda+1/2\left(1+1/\Lambda\right)\left(\partial_{p_\ast}-\partial_{p}\right)\Delta\left(p,p_\ast\right)\right|^{-1}.
\end{eqnarray}
The significant parameter in the opinion formation is the temperature ($\theta$) in the closed opinion system, because $\theta \rightarrow \infty$ means that $|m|$ of all the peoples approximate to unity, where $|m|=1$ corresponds to the complete agreement or disagreement on the single issue, namely, complete decision making. In our relativistic kinetic model, we never postulate a massless particle. Therefore, the people with the complete decision making, namely, $|m|=1$, is not considered. We, however, have a question, \textit{What is the temperature in the closed opinion system ?} The possible answer to this question is that the temperature in the closed opinion system is equivalent to the global interest in the single issue. \textcolor{red}{The increase of the interest yields the increase of the amplitudes of the self-thinking, which corresponds to the amplitude of thermal fluctuations of opinions. Therefore, we can consider that the strength of the interest is equivalent to the amplitude of the self-thinking (thermal fluctuations), which is related to the temperature.} Provided that all the peoples have high interests in the single issue, $|m|$ of all the peoples approximate to unity, namely, complete decision making. \textcolor{red}{For instance, $\theta=\infty$ means that the number of peoples with the complete agreement ($m=1$) is equal to that with the complete disagreement ($m=-1$), when the mean opinion is neutral, namely, $\bar{m}=0$}. Consequently, we conjecture that the global interest in the single issue decreases by the binary inelastic collision without randomly perturbed motion (self-thinking), whereas the global interest increases by the self-thinking, namely, randomly perturbed motion $\Delta$.
\subsection{Characteristics of relativistic inelastic-collisions with randomly perturbed motion}
In this subsection, we investigate the characteristics of relativistic inelastic-collisions with randomly perturbed motion. Therefore, we set $B=0$ in Eq. (5) to neglect effects of the external force by the second (or political) party.\\ 
The temporal evolution of the flow velocity $N^\alpha(=\int_{-\infty}^\infty p^\alpha f dp/p^0$) is obtained by multiplying $p^\alpha /p^0$ by both sides of Eq. (5) and integrating over the momentum space $dp/p^0$ as
\begin{eqnarray}
d_t N^{\alpha}&=&\frac{A}{2}\int_{-\infty}^{\infty}\int_{-\infty}^{\infty} \left(\frac{{p^\alpha}^\prime}{{p^0}^\prime}+\frac{{p^\alpha_\ast}^\prime}{{p^0_\ast}^\prime}-\frac{{p^\alpha}}{p^0}-\frac{{p^\alpha_\ast}}{p^0_\ast}\right) f\left(t,p\right) f\left(t,p_\ast\right) F \frac{d p_\ast}{p^0_\ast} \frac{d p}{p^0}.
\end{eqnarray}
In the right hand side of Eq. (10), $\int_{-\infty}^{\infty}\int_{-\infty}^{\infty} \left({p^\alpha}^\prime/{p^0}^\prime+{p^\alpha_\ast}^\prime/{p^0_\ast}^\prime-p^\alpha/p^0-p^\alpha_\ast/p^0_\ast \right) f\left(t,p\right) f\left(t,p_\ast\right) F \frac{dp_\ast}{p^0_\ast} \frac{dp}{p^0}=0$, when $\alpha=0$. Here, we must remind that the number density ($n$) temporally changes in accordance with the change of the mean opinion, $\bar{m}$, owing to $d_t N^0=d_t(n U^0)=0$, where $U^\alpha$ is two-vectors, that is defined by $U^\alpha=\gamma\left(\bar{m}\right)\left(1,\bar{m}\right)$. On the other hand, we obtain $d_t N^1 \neq 0$, because we obtain ${p^1}^\prime/{p^0}^\prime+{p^1_\ast}^\prime/{p^0_\ast}^\prime-p^1/p^0-p^1_\ast/p^0_\ast=m^\prime+m_\ast^\prime-m-m_\ast \neq 0$ from Eqs. (6) and (7).\\ 
The temporal evolution of the energy-momentum tensor, $T^{\alpha\beta}$ ($=\int_{-\infty}^\infty  p^\alpha p^\beta f dp/p^0$), is obtained by multiplying $p^\alpha p^\beta/p^0$ by both sides of Eq. (5) and integrating over the momentum space $dp/p^0$ as
\begin{eqnarray}
d_t T^{\alpha\beta}&=&\frac{A}{2}\int_{-\infty}^{\infty}\int_{-\infty}^{\infty} \left(\frac{{p^\alpha}^\prime{p^\beta}^\prime}{{p^0}^\prime}+\frac{{p^\alpha_\ast}^\prime {p^\beta_\ast}^\prime}{{p^0_\ast}^\prime}-\frac{{p^\alpha}{p^\beta}}{p^0}-\frac{{p^\alpha_\ast}{p^\beta_\ast}}{p^0_\ast}\right) f\left(t,p\right) f\left(t,p_\ast\right) F \frac{d p_\ast}{p^0_\ast} \frac{d p}{p^0}.
\end{eqnarray}
In the right hand side of Eq. (11),\\
$(A/2)\int_{-\infty}^{\infty}\int_{-\infty}^{\infty} \left({p^\alpha}^\prime{p^\beta}^\prime/{p^0}^\prime+{p^\alpha_\ast}^\prime{p^\beta_\ast}^\prime/{p^0_\ast}^\prime-p^\alpha p^\beta/p^0-p^\alpha_\ast p^\beta_\ast/p^0_\ast \right) f\left(t,p\right) f\left(t,p_\ast\right) F \frac{dp_\ast}{p^0_\ast} \frac{dp}{p^0}$ is equal to 0 from Eqs. (6) and (7), when $\alpha=0$ and $\beta=1$.\\
Consequently, the temporal evolution of $T^{01}$ is obtained using Eq. (A4), when $f=f_{MJ}=n/(2K_1(\chi))\exp(-\chi p^\alpha U_\alpha)$ ($f_{MJ}$: one dimensional Maxwell-J$\ddot{\mbox{u}}$ttner function \cite{Kremer}), as
\begin{eqnarray}
d_t T_E^{01}&=&d_t \left(n \frac{K_2(\chi)}{K_1(\chi)}U^0 U^1\right)\nonumber \\
&=&n U^0 d_t \left(U^1 \frac{K_2(\chi)}{K_1(\chi)}\right)=0,
\end{eqnarray}
where we used $d_t N^0=d_t \left(n U^0\right)=0$.\\
From Eq. (12), we obtain
\begin{eqnarray}
U^1(t)=\mathcal{C} \frac{K_1(\chi(t))}{K_2(\chi(t))},
\end{eqnarray}
where $\mathcal{C}=U^1(0) \frac{K_2(\chi(0))}{K_1(\chi(0))}$.\\ $K_1(\chi(t))/K_2(\chi(t))$ decreases, as $\chi$ decreases. As a result, $|\bar{m}|$ increases, as $\chi$ ($\theta$) increases (decreases), whereas $|\bar{m}|$ decreases, as $\chi$ $(\theta)$ decreases (increases). Similarly, $\chi$ ($\theta$) increases (decreases), as $|\bar{m}|$ increases, whereas $\chi$ $(\theta)$ decreases (increases), as $|\bar{m}|$ decreases. Of course, $\bar{m}(t)=0$, when $\bar{m}(0)=0$. \\\\
\textbf{Remark 1}\\\\
\textit{The mean opinion $\bar{m}$ never be conserved by relativistic inelastic-collisions, even when the total momentum of binary colliding opinions is conserved. Provided that $f=f_{MJ}$, $|\bar{m}|$ is inversely proportion to the global interest and its vice versa. $\lim_{t \rightarrow \infty} U^1(t)=\mathcal{C}$ indicates that $U^1$ converges to not $U^1(0)$ but $\mathcal{C}$, because $\lim _{t \rightarrow \infty} {K_1(\chi(t))}/{K_2(\chi(t))}=1$ is obtained by $\lim _{t \rightarrow \infty} \chi(t)=\infty$ owing to relativistic inelastic-collisions.}\\\\
Next, we consider the heating via the self-thinking by incorporating the randomly perturbed motion into the binary inelastic collisions. Here, we restrict ourselves to the binary elastic collisions to focus on the heating process via the self-thinking.\\ Firstly, from the definition of the energy, $E=U_\alpha U_\beta T^{\alpha\beta}$, we obtain following relation by multiplying $U^\alpha U^\beta$ by both sides of Eq. (11),
\begin{eqnarray}
d_t E-T^{\alpha\beta}d_t \left(U_\alpha U_\beta\right) &=&U_\alpha U_\alpha\frac{A}{2}\int_{-\infty}^{\infty}\int_{-\infty}^{\infty} \left(\frac{{p^\alpha}^\prime{p^\alpha}^\prime}{{p^0}^\prime}+\frac{{p^\alpha_\ast}^\prime {p^\alpha_\ast}^\prime}{{p^0_\ast}^\prime}-\frac{{p^\alpha}{p^\alpha}}{p^0}-\frac{{p^\alpha_\ast}{p^\alpha_\ast}}{p^0_\ast}\right) f(t,p) f(t,p_\ast) F \frac{dp_\ast}{p^0_\ast} \frac{dp}{p^0}. \nonumber \\
\end{eqnarray}
Provided $\bar{m}(t)=0$ for $0 \le t$, we can rewrite Eq. (14) as
\begin{eqnarray}
d_t E&=&\frac{A}{2}\int_{-\infty}^{\infty}\int_{-\infty}^{\infty} \left({p^0}^\prime+{p^0_\ast}^\prime-p^0-p^0_\ast \right) f(t,p) f(t,p_\ast) F \frac{dp_\ast}{p^0_\ast} \frac{dp}{p^0}.
\end{eqnarray}
In the right hand side of Eq. (15), $\left({p^0}^\prime+{p^0_\ast}^\prime-p^0-p^0_\ast \right)$ has both positive and negative value, when $\Lambda=1$, whereas $\left({p^0}^\prime+{p^0_\ast}^\prime-p^0-p^0_\ast \right)$ always has the positive value under the nonrelativistic limit when $\Lambda=1$, because
\begin{eqnarray}
&&{p^0}^\prime+{p^0_\ast}^\prime-p^0-p^0_\ast=\sqrt{1+{p^\prime}^2}+\sqrt{1+{p_\ast^\prime}^2}-\sqrt{1+p^2}-\sqrt{1+p_\ast^2} \nonumber \\
&& \simeq \frac{1}{2}{p^\prime}^2+\frac{1}{2}{p^\prime_\ast}^2-\frac{1}{2}p^2-\frac{1}{2}p_\ast^2=\Delta^2 \ge 0~~\because~~p^\prime\ll 1~~\wedge~~p^\prime_\ast \ll 1,~~\wedge~~p \ll 1~~\wedge~~p_\ast \ll 1. \nonumber\\
\end{eqnarray}
The right hand side of Eq. (15) is rewritten by assuming $f=f_{MJ}$, when $\bar{m}=0$ and $\Lambda=1$, as
\begin{eqnarray} 
&&\frac{n^2 A}{8K_1^2(\chi)}\int_{-\frac{\pi}{2}}^{\frac{\pi}{2}} d \textcolor{red}{\varpi} \int_{-\frac{\pi}{2}}^{\frac{\pi}{2}} d \sigma 
\left|\sin\textcolor{red}{\varpi}-\sin \sigma\right| e^{-\chi(\frac{1}{\cos\sigma}+\frac{1}{\cos\textcolor{red}{\varpi}})}\nonumber \\
&&\times \left(\sqrt{(\tan\textcolor{red}{\varpi}-\Delta)^2+1}+\sqrt{(\Delta+\tan\sigma)^2+1}-\frac{1}{\cos\textcolor{red}{\varpi}}-\frac{1}{\cos\sigma}\right),
\end{eqnarray}
where we used $p=\tan \textcolor{red}{\varpi}$ and $p_\ast=\tan \sigma$.\\
The positivity of Eq. (17) can be proved, when $\left|\Delta\right| \ll 1$, as
\begin{eqnarray}
&&\frac{n^2 A}{8 K_1^2(\chi)}\int_{-\frac{\pi}{2}}^{\frac{\pi}{2}} d \textcolor{red}{\varpi} \int_{-\frac{\pi}{2}}^{\frac{\pi}{2}} d \sigma 
\left|\sin\textcolor{red}{\varpi}-\sin \sigma\right| e^{-\chi(\frac{1}{\cos\sigma}+\frac{1}{\cos\textcolor{red}{\varpi}})}\nonumber \\
&&\left(\frac{1}{\cos\textcolor{red}{\varpi}}\sqrt{1-2\Delta \sin\textcolor{red}{\varpi}\cos\textcolor{red}{\varpi}+\cos^2 \textcolor{red}{\varpi} \Delta^2}+\frac{1}{\cos\sigma}\sqrt{1+2\Delta \sin \sigma \cos \sigma+\cos^2 \sigma\Delta^2}-\frac{1}{\cos\textcolor{red}{\varpi}}-\frac{1}{\cos\sigma}\right), \nonumber \\
&\simeq& \frac{n^2 A}{8 K_1^2(\chi)}\int_{-\frac{\pi}{2}}^{\frac{\pi}{2}} d \textcolor{red}{\varpi} \int_{-\frac{\pi}{2}}^{\frac{\pi}{2}} d \sigma \left|\sin\textcolor{red}{\varpi}-\sin \sigma\right| e^{-\chi(\frac{1}{\cos\sigma}+\frac{1}{\cos\textcolor{red}{\varpi}})} \left[\Delta (\sin\sigma-\sin \textcolor{red}{\varpi})+\frac{1}{2}\Delta^2\left(\frac{1}{\cos\textcolor{red}{\varpi}}+\frac{1}{\cos\sigma}\right)\right] \nonumber \\
&=&\frac{n^2 A}{16 K_1^2(\chi)}\int_{-\frac{\pi}{2}}^{\frac{\pi}{2}} d \textcolor{red}{\varpi} \int_{-\frac{\pi}{2}}^{\frac{\pi}{2}} d \sigma \left|\sin\textcolor{red}{\varpi}-\sin \sigma\right| e^{-\chi(\frac{1}{\cos\sigma}+\frac{1}{\cos\textcolor{red}{\varpi}})} \Delta^2 \left(\frac{1}{\cos\textcolor{red}{\varpi}}+\frac{1}{\cos\sigma}\right)  \ge 0,
\end{eqnarray}
where we used the relation $\int_{-\pi/2}^{\pi/2} d \textcolor{red}{\varpi} \int_{-\pi/2}^{\pi/2} d \sigma \Delta \left|\sin\textcolor{red}{\varpi}-\sin \sigma\right| e^{-\chi(\frac{1}{\cos\sigma}+\frac{1}{\cos\textcolor{red}{\varpi}})}\left(\sin\sigma-\sin \textcolor{red}{\varpi}\right)=0$, because $\sigma \left|\sin\textcolor{red}{\varpi}-\sin \sigma\right| e^{-\chi(\frac{1}{\cos\sigma}+\frac{1}{\cos\textcolor{red}{\varpi}})}\left(\sin\sigma-\sin \textcolor{red}{\varpi}\right)$ is the odd function. Meanwhile, the direct evaluation of Eq. (17) for the proof of the positivity of Eq. (17) is not obtained owing to its mathematical difficulties. From later numerical results, we confirm that randomly perturbed motion sometimes yields the cooling of $\theta$.\\
Next, we consider the cooling process by relativistic inelastic-collisions, when the self-thinking term is neglected, namely, $\Delta=0$. Here, we consider the temporal evolution of $T^{0\alpha\beta}$ instead of that of $T^{\alpha\beta}$ to remove terms divided by $p^0$ in Eq. (11). Additionally, we assume that the distribution function is expressed by Maxwell-J$\ddot{\mbox{u}}$ttner function, namely, $f=f_{MJ}$.\\
The temporal evolution of $T^{0\alpha \beta}_E=\int_{-\infty}^\infty p^0 p^\alpha p^\beta f_{MJ} dp/p^0$ is written, when $\Delta=0$ and $B=0$, as
\begin{eqnarray}
d_t T^{0\alpha \beta}_E=\frac{A}{2}\int_{-\infty}^{\infty}\int_{-\infty}^{\infty} \left({p^\alpha}^\prime{p^\beta}^\prime+{p^\alpha_\ast}^\prime {p^\beta_\ast}-p^\alpha p^\beta-p^\alpha_\ast p^\beta_\ast \right) f_{MJ}\left(t,p\right) f_{MJ}\left(t,p_\ast\right) F \frac{dp_\ast}{p^0_\ast} \frac{d p}{p^0},
\end{eqnarray}
We introduce two vectors $P^\alpha$ and $Q^\alpha$, which are defined by
\begin{eqnarray}
&&P^\alpha=p^\alpha+p^\alpha_\ast,~~~~{P^{\alpha}}^\prime={p^\alpha}^\prime+{p^\alpha_\ast}^\prime, \nonumber \\
&&Q^\alpha=p^\alpha-p^\alpha_\ast,~~~~{Q^{\alpha}}^\prime={p^\alpha}^\prime-{p^\alpha_\ast}^\prime,
\end{eqnarray}
where we remind that $p^1+p^1_\ast={p^1}^\prime+{p^1_\ast}^\prime$ and $p^0+p^0_\ast \ge {p^0}^\prime+{p^0_\ast}^\prime$.\\
We obtain following relations from Eq. (20)
\begin{eqnarray}
&&P^\alpha Q_\alpha=0,~~~{P^\alpha}^\prime {Q_\alpha}^\prime=0, \\
&&P^2=P^\alpha P_\alpha=4+Q^\alpha Q_\alpha=4+Q^2.
\end{eqnarray}
From inverse transformation of Eq. (20), we obtain
\begin{eqnarray}
p^\alpha&=&\frac{1}{2}\left(P^\alpha+Q^\alpha\right),~~~p^\alpha_\ast=\frac{1}{2}\left(P^\alpha-Q^\alpha\right),\nonumber \\
{p^\alpha}^\prime&=&\frac{1}{2}\left({P^\alpha}^\prime+{Q^\alpha}^\prime\right),~~~{p^\alpha_\ast}^\prime=\frac{1}{2}\left({P^\alpha}^\prime-{Q^\alpha}^\prime\right),
\end{eqnarray}
Substituting Eq. (23) into Eq. (19), we obtain
\begin{eqnarray}
&&d_t T^{0\alpha \beta}_E= \nonumber \\
&&\frac{A}{8}\int_{-\infty}^{\infty}\int_{-\infty}^{\infty} \left[\left({P^\alpha}^\prime{P^\beta}^\prime-P^\alpha P^\beta \right)+\left( {Q^\alpha}^\prime{Q^\beta}^\prime-Q^\alpha Q^\beta \right)\right] f_{MJ}\left(t,p\right) f_{MJ}\left(t,p_\ast\right) g_{\o} dP dQ,
\end{eqnarray}
where we used the relation $F dp/p_0 dp_\ast/{p_0}_\ast=1/2 g_{\o} dPdQ$ \cite{Cercignani}.\\
The integration of ${A}/{8}\int_{-\infty}^{\infty}\int_{-\infty}^{\infty} \left({P^\alpha}^\prime{P^\beta}^\prime-P^\alpha P^\beta \right) f_{MJ}\left(t,p\right) f_{MJ}\left(t,p_\ast\right) g_{\o} dP dQ$ in the right hand side of Eq. (24) is markedly difficult, whereas ${P^1}^\prime{P^1}^\prime-P^1 P^1=0$.\\
Hereafter, we can neglect the integration ${A}/{8}\int_{-\infty}^{\infty}\int_{-\infty}^{\infty} \left({P^\alpha}^\prime{P^\beta}^\prime-P^\alpha P^\beta \right) f_{MJ}\left(t,p\right) f_{MJ}\left(t,p_\ast\right) g_{\o} dP dQ$ with $\alpha=0$ or $\beta=0$ in the right hand side of Eq. (24), because we will investigate the temporal evolution of $T^{011}$ in later discussion.\\
The center of mass system yields relations using Eqs. (6) and (7) 
\begin{eqnarray}
P^\alpha=\left(P^0,0\right),~~~Q^\alpha=\left(0,Q\right),~~~{Q^\alpha}^\prime=-\Lambda\left(0,Q\right).
\end{eqnarray}
From Eq. (25), we obtain
\begin{eqnarray}
{Q^\alpha}^\prime{Q^\beta}^\prime-Q^\alpha Q^\beta=-Q^2\left(1-\Lambda^2\right)
\left(
\begin{array}{cc}
0 & 0 \\
0 & 1 
\end{array}
\right)=-Q^2\left(1-\Lambda^2\right)\left(\frac{P^\alpha P^\beta}{P^2}-\eta^{\alpha \beta} \right),
\end{eqnarray}
where $\eta^{\alpha\beta}=\mbox{diag}(1,-1,-1,-1)$.\\
M\o ller's relative velocity in the center of mass system is \cite{Cercignani}
\begin{eqnarray}
g_{\o}=2 \frac{Q}{P^0}.
\end{eqnarray}
From Eqs. (26) and (27), we rewrite Eq. (24), when $\alpha \neq 0 \wedge \beta \neq 0$, as
\begin{eqnarray}
&& d_t T_E^{0\alpha\beta}\nonumber \\
&=&\frac{A}{8}\int_{-\infty}^{\infty}\int_{-\infty}^{\infty} \left({Q^\alpha}^\prime{Q^\beta}^\prime-Q^\alpha Q^\beta \right) f_{MJ}\left(t,p\right) f_{MJ}\left(t,p_\ast\right) g_{\o} dP dQ \nonumber \\
&=&-\left(1-\Lambda^2\right)\frac{A}{4}\int_{-\infty}^{\infty}\int_{-\infty}^{\infty} \left(\frac{P^\alpha P^\beta}{P^2}-\eta^{\alpha \beta} \right) Q^3 f_{MJ}\left(t,p\right) f_{MJ}\left(t,p_\ast\right) \frac{dP}{P^0}dQ \nonumber \\
&=&-\left(1-\Lambda^2\right) n^2 \frac{A}{16 K_1(\chi)^2}\int_{-\infty}^{\infty} \left(\frac{{Z^\star}^{\alpha\beta}}{4+Q^2}-\eta^{\alpha \beta} Z^\star \right) Q^3 dQ\nonumber \\
&=&-\left(1-\Lambda^2\right) n^2 \frac{A}{4 K_1(\chi)^2} \int_{2}^{\infty} \left(K_2\left(\chi Q^\star\right) U^\alpha U^\beta Q^\star-\eta^{\alpha\beta} \frac{K_1\left(\chi Q^\star\right)}{\chi}-\eta^{\alpha \beta} K_0 \left(\chi Q^\star\right) Q^\star \right) \left({Q^\star}^2-4\right) d Q^\star \nonumber \\
&=&-\left(1-\Lambda^2\right) n^2 \frac{A}{4 \chi K_1(\chi)^2} \int_{2\chi}^{\infty} \left(K_2\left(x\right) U^\alpha U^\beta \frac{x}{\chi}-\eta^{\alpha\beta} \frac{K_1\left(x\right)}{\chi}-\eta^{\alpha \beta} K_0 \left(\chi Q^\star\right) \frac{x}{\chi} \right) \left(\frac{x^2}{\chi^2}-4\right) d x \nonumber \\
&=&-\left(1-\Lambda^2\right) n^2 \frac{A}{\chi^3 K_1(\chi)^2} \left[2\chi K_2\left(2\chi\right) U^\alpha U^\beta-\eta^{\alpha \beta} \left(2 \chi K_0(2\chi)+K_1(2\chi)\right)\right],
\end{eqnarray}
where $Z^\star$ and ${Z^\star}^{\alpha\beta}$ are defined in Eqs. (A8) and (A9), $Q^\star=\sqrt{Q^2+4}$ and $x=Q^\star/\chi$.\\
From Eqs. (12) and (28), we obtain the temporal evolution of $T^{011}_E$ as
\begin{eqnarray}
d_t T_E^{011}&=&-\left(1-\Lambda^2\right) n^2 \frac{A}{\chi^3 K_1(\chi){}^2} \left[2\chi K_2\left(2\chi\right) \left(U^1\right)^2 -\eta^{11} \left(2 \chi K_0(2\chi)+K_1(2\chi)\right)\right],\nonumber \\
&=& -\left(1-\Lambda^2\right) n^2 \frac{A}{\chi^3 K_1(\chi){}^2} \left[2\chi K_2\left(2\chi\right) \left(\mathcal{C}\frac{K_1(\chi)}{K_2(\chi)}\right)^2 -\eta^{11} \left(2 \chi K_0(2\chi)+K_1(2\chi)\right)\right].
\end{eqnarray}
We obtain following relation using Eqs. (A7) and (12)
\begin{eqnarray}
d_t T^{011}_E&=&\frac{n}{\sqrt{1+\mathcal{C}^2 K_1(\chi)^2/K_2(\chi)^2}} \nonumber \\
&&\left[2 \chi ^2 \left(\mathcal{C}^2 \chi^4+8\mathcal{C}^2 \chi^2+3\chi^2+8 \right) K_1(\chi ){}^3 K_0(\chi ){}^2+2 \chi  \left(6\chi ^2-\mathcal{C}^2 \chi ^2 \left(\chi^2-8\right)+24\right) K_1(\chi ){}^4 K_0(\chi) \right. \nonumber \\
&&\left. -2 \left(\mathcal{C}^2 \chi^4 \left(\chi ^2+6\right)-4 \chi^2-16\right) K_1(\chi ){}^5+\chi ^3 \left(4 \mathcal{C}^2\chi ^2+\chi ^2-8\right) K_1(\chi ){}^2 K_0(\chi ){}^3 \right. \nonumber \\
&& \left. -\chi ^5 K_0(\chi){}^5-6 \chi ^4 K_1(\chi ) K_0(\chi ){}^4\right]\left(\chi ^4 K_1(\chi ){}^2 K_2(\chi){}^3\right)^{-1} \frac{d \chi^{-1}}{d t},
\end{eqnarray}
where we used $d_t \left(n U^0\right)=0$.\\
From Eqs. (29) and (30), we obtain
\begin{eqnarray}
d_t \chi^{-1}&=&-n A \left(1-\Lambda^2\right) \psi_1\left(\chi,\mathcal{C}\right) \chi^{-1}, \nonumber \\
\psi_1\left(\chi,\mathcal{C}\right) &=&\left(1+\mathcal{C}^2 K_1(\chi)^2/K_2(\chi)^2 \right)^{-\frac{1}{2}} \left[\chi^2 K_2(\chi) \left(2 \mathcal{C}^2 \chi K_2(2 \chi ) K_1(\chi ){}^2+(2 \chi  K_0(2 \chi )+K_1(2 \chi )) K_2(\chi ){}^2\right)\right] \nonumber \\
&& \left[2 \chi ^2 \left(\mathcal{C}^2 \chi^4+8 \mathcal{C}^2 \chi^2+3\chi^2+8\right) K_1(\chi ){}^3 K_0(\chi ){}^2+2 \chi \left(6\chi ^2-\chi ^4 \mathcal{C}^2+8 \chi ^2 \mathcal{C}^2+24\right) K_1(\chi){}^4 K_0(\chi) \right. \nonumber \\
&& \left. -2 \left(\mathcal{C}^2 \chi^6+6 \mathcal{C}^2 \chi ^4-4\chi ^2-16\right) K_1(\chi ){}^5+\chi ^3 \left(4 \mathcal{C}^2 \chi ^2+\chi ^2-8\right) K_1(\chi ){}^2 K_0(\chi ){}^3 \right. \nonumber \\
&& \left. -\chi ^5 K_0(\chi){}^5-6 \chi ^4 K_1(\chi ) K_0(\chi ){}^4\right]^{-1}.
\end{eqnarray}
From Eq. (31), we find that the cooling rate parameter $\psi_1\left(\chi,\mathcal{C}\right)$ depends on the frame via $\bar{m}(0)$.
\begin{center}
\includegraphics[width=0.9\textwidth]{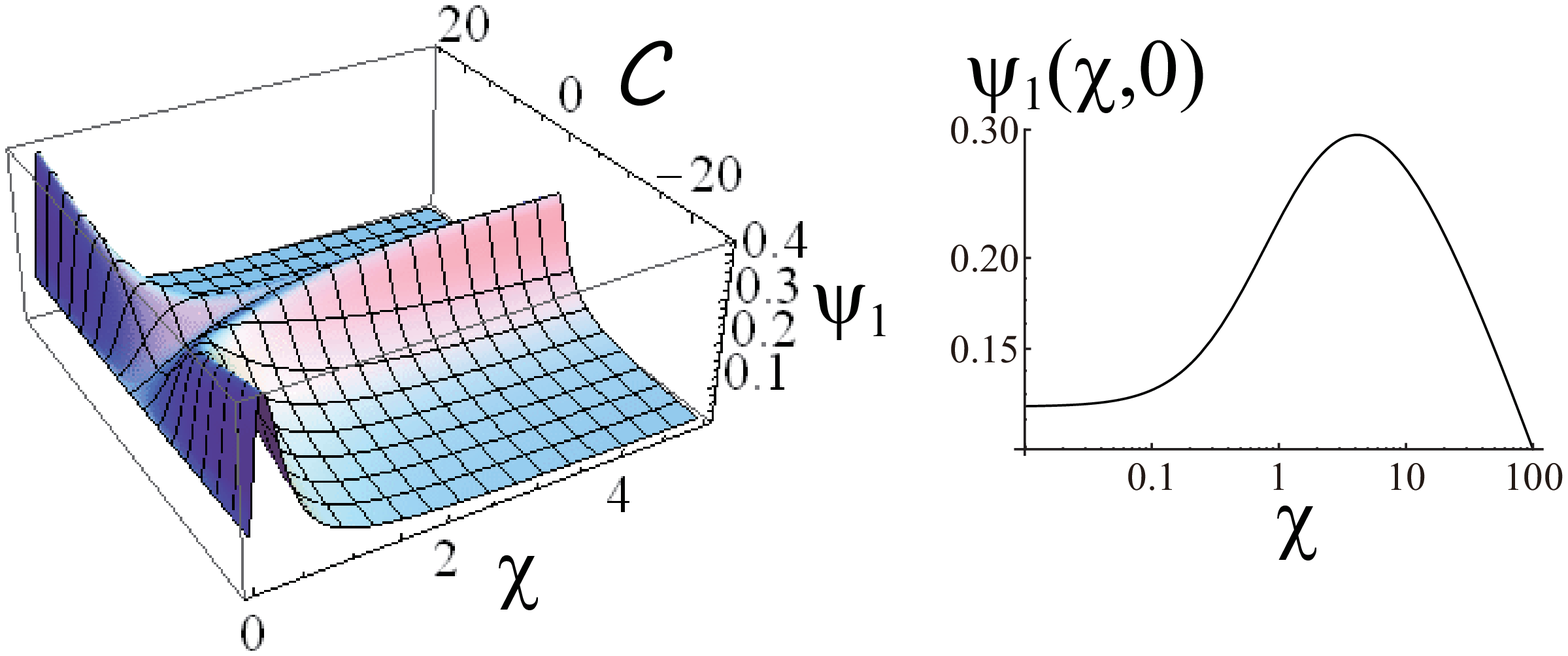}\\
\footnotesize{FIG. 1: $\psi_1\left(\chi,\mathcal{C}\right)$ versus $\chi$ and $\mathcal{C}$ (left frame), and $\psi_1\left(\chi,0\right)$ versus $\chi$ (right frame).}
\end{center}
The left frame of Fig. 1 shows $\psi_1\left(\chi,\mathcal{C}\right)$ versus $\chi$ and $\mathcal{C}$. $\psi_1\left(\chi,\mathcal{C}\right)$ is symmetric at both sides of $\mathcal{C}=0$. $\psi_1\left(\chi,\mathcal{C}\right)$ approximates to 0, when $\chi$ approximates to $\infty$, whereas $\psi_1\left(\chi,\mathcal{C}\right)$ approximates to $1/8$, when $\chi$ approximates to 0. Additionally, $\psi_1\left(\chi,\mathcal{C}\right)$ has a peak, which moves toward to $\chi=0$, as $|\mathcal{C}|$ increases.\\
Provided that $\mathcal{C}=0$ ($\bar{m}(0)=0$), we obtain $\psi_1\left(\chi,0\right)$ from Eq. (31) as
\begin{eqnarray}
\psi_1\left(\chi,0\right)=\frac{2 \chi  K_0\left(2 \chi\right)+K_1\left(2 \chi\right)}{\chi \left(\chi ^2+4\right) K_1\left(\chi\right){}^2-\chi ^3 K_0\left(\chi\right){}^2}.
\end{eqnarray}
The right frame of Fig. 1 shows $\psi_1\left(\chi,0\right)$ versus $\chi$. $\psi_1\left(\chi,0\right)$ has its peak value $0.295$ at $\chi \simeq 4.14$, whereas we obtain $\lim_{\chi \rightarrow 0} \psi_1(\chi,0)=1/8$ and $\lim_{\chi \rightarrow \infty} \psi_1(\chi,0)=2/\sqrt{\pi \chi}$, which coincides with the cooling rate parameter, which is calculated using the one dimensional nonrelativistic inelastic-Boltzmann equation.
\subsection{Effects of external force by the second party on opinion formation}
We consider the temporal evolution of $\partial_0 T_E^{0\alpha\beta\gamma\delta...}$ ($T_E^{0\alpha\beta\gamma\delta...}=\int p^0 p^\alpha p^\beta p^\gamma p^\delta ... f_{MJ} dp/p^0$) via the external force by the second (or political) party, which is defined as
\begin{eqnarray}
d_t T_E^{0\alpha\beta\gamma\delta...}&=&B \int_{-\infty}^{\infty} p^0 p^\alpha p^\beta p^\gamma p^\delta ... \frac{\partial \left(p-P\right) f_{MJ}}{\partial p}\frac{d p}{p^0}, \nonumber \\
&=&\frac{n B}{2 K_1(\chi)}\left[Z^{0\alpha\beta\gamma\delta...}-\chi\left(Z^{11\alpha\beta\gamma\delta...}U^0-Z^{01\alpha\beta\gamma\delta...}U^1\right)+P\chi\left(Z^{1\alpha\beta\gamma\delta...}U^0-Z^{0\alpha\beta\gamma\delta...}U^1\right)\right], \nonumber \\
\end{eqnarray}
where $Z^{\alpha\beta\gamma}$ and $Z^{\alpha\beta\gamma\delta}$ are defined in Eqs. (A7) and (A8).\\
From Eq. (33), we readily obtain $d_t T_E^0=d_t N_E^0=d_t\left(nU^0\right)=0$.\\
Similarly. we obtain $d_t T_E^{01}$ and $d_t T_E^{011}$ as
\begin{eqnarray}
d_t T_E^{01}&=&-B n U^0 \left(\frac{K_2(\chi)}{K_1(\chi)} U^1-P \right),\\
d_t T_E^{011}&=& -2 B n U^0 \left(\frac{K_3(\chi)}{K_1(\chi)} \left(U^1\right)^2+\frac{1}{\chi}\frac{K_2(\chi)}{K_1(\chi)}-P \frac{K_2(\chi)}{K_1(\chi)} U^1\right),
\end{eqnarray}
Using the relation $d_t \left(n U^0\right)=0$ and Eqs. (A6) and (A7), we can rewrite Eqs. (34) and (35), respectively, as
\begin{eqnarray}
&&d_t \left(\frac{K_2(\chi)}{K_1(\chi)} U^1\right)=-B\left(\frac{K_2(\chi)}{K_1(\chi)} U^1-P \right),\\
&&d_t \left(\frac{K_3(\chi)}{K_1(\chi)} (U^1)^2+\frac{1}{\chi}\frac{K_2(\chi)}{K_1(\chi)}\right)=-2 B \left(\frac{K_3(\chi)}{K_1(\chi)} \left(U^1\right)^2+\frac{1}{\chi}\frac{K_2(\chi)}{K_1(\chi)}-P \frac{K_2(\chi)}{K_1(\chi)} U^1\right)
\end{eqnarray}
From Eq. (36), we readily obtain
\begin{eqnarray}
U^1(t)=\frac{K_1(\chi(t))}{K_2(\chi(t))}\left[P+\mathscr{C} \exp(-B t)\right],
\end{eqnarray}
where $\mathscr{C}=\left(K_2(\chi)/K_1(\chi)(U^1)-P\right)_{t=0}$.\\
Substituting Eq. (38) into Eq. (37), we cannot obtain general solution of $\chi(t)$ owing to mathematical difficulties. Then, we restrict ourselves to $1 \ll t$, which allows us to assume $U^1=K_1(\chi)/K_2(\chi) P$ in Eq. (38).\\
Substituting $U^1=K_1(\chi)/K_2(\chi) P$ into Eq. (37), we obtain 
\begin{eqnarray}
d_t \chi^{-1}&=&-B \psi_2\left(\chi,P\right) \chi^{-1}, \nonumber \\
\psi_2\left(\chi,P\right)&=& -\left[4 \chi K_1(\chi ) K_2(\chi ) \left(P^2 \chi  K_1(\chi ){}^3+4 P^2 K_2(\chi ) K_1(\chi ){}^2-P^2 \chi  K_2(\chi ){}^2 K_1(\chi )+K_2(\chi ){}^3\right)\right] \nonumber \\
&&\left[4 P^2 K_1(\chi ){}^2 \left\{-2 \chi ^2 K_0(\chi ){}^3-\chi  \left(\chi ^2+8\right) K_1(\chi ) K_0(\chi ){}^2+\left(\chi ^2-8\right) K_1(\chi ){}^2 K_0(\chi) \right. \right. \nonumber \\
&& \left. \left. +\chi \left(\chi ^2+6\right) K_1(\chi ){}^3\right\}+\chi ^2 K_2(\chi ){}^5-2 \chi ^2 K_1(\chi ){}^2 K_2(\chi ){}^3+\chi  (\chi  K_0(\chi )-6 K_1(\chi )) K_2(\chi ){}^4\right]^{-1}. \nonumber \\
\end{eqnarray}
Figure 2 shows $\psi_2\left(\chi,P\right)$ versus $\chi$ and $P$. The peak of $\psi_2\left(\chi,P\right)$ increases, when $|P|$ increases, as shown in Fig. 2. Additionally, $\chi$, which yields two peaks of $\psi_2$ in positive and negative $P$, increases, as $|P|$ increases. $\psi_2\left(\chi,P\right)$ approximates to unity under $\chi \rightarrow 0$, whereas $\psi_2\left(\chi,P\right)$ approximates to 2 under $\chi \rightarrow \infty$, which is equivalent to $\psi_2$ for the nonrelativistic gas \cite{Santos}.
\begin{center}
\includegraphics[width=0.5\textwidth]{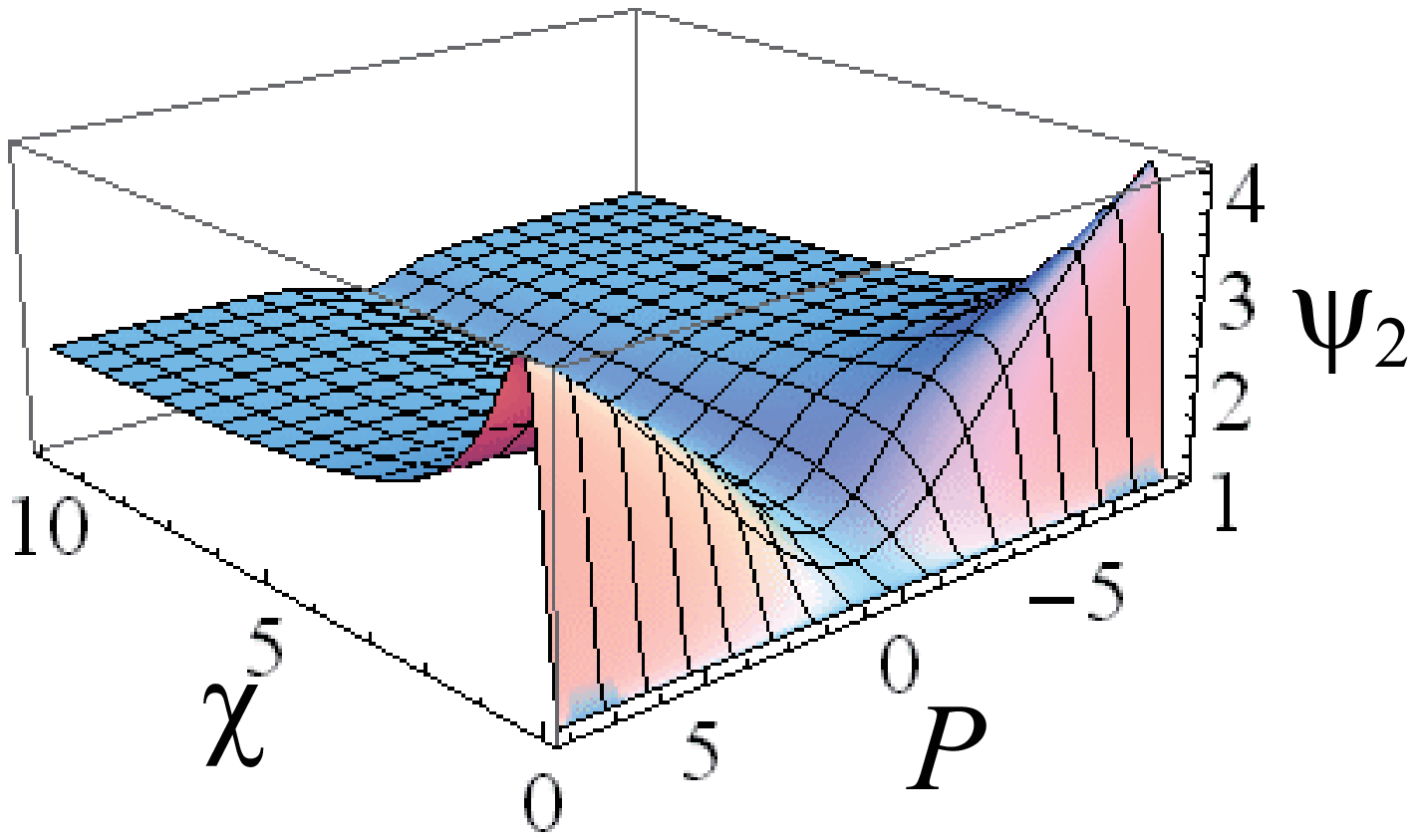}\\
\footnotesize{FIG. 2: $\psi_2$ versus $\chi$ and $P$.}
\end{center}
Finally, we must remind that the cooling rate of $\theta$ via relativistic inelastic-collisions or external force by the second ( or political) party is not always described by $\psi_1(\chi,\mathcal{C})$ in Eq. (31), or $\psi_2\left(\chi,P\right)$ in Eq. (39), because the distribution function never approximates to the equilibrium distribution function, namely, Maxwell-J$\ddot{\mbox{u}}$ttner function, because H theorem is not guaranteed by Eq. (5). Therefore, we must calculate the cooling rate on the basis of the nonequilibrium distribution function, such as Grad's N-moment equation \cite {Israel} to consider nonequilibrium effects on the cooling rate of $\theta$. Meanwhile, such a calculation of the cooling rate on the basis of Grad's N-moment equation is set as our future study.
\section{Numerical study of relativistic kinetic model}
In this paper, we investigate the opinion formation, which is described by the Eq. (5) by changing parameters, $\Lambda$, $\Delta$ and $B$ in Eq. (5), whereas $A$ in Eq. (5) is fixed to the constant value, namely, $A=1$. Additionally, physical quantities such as the density, mean opinion ($\bar{m}$), and global interest ($\theta$) are calculated using Eckart's decomposition of $N^\alpha=\int_{-\infty}^\infty p^\alpha f dp/p^0$ and $T^{\alpha\beta}=\int_{-\infty}^\infty p^\alpha p^\beta f dp/p^0$ \cite{Eckart}. Consequently, $n$, $\Pi^{\left<\alpha \beta\right>}$, $p+\Pi$ (static pressure+dynamic pressure), $q^\alpha$ (heat flux), and $e=1/\chi+K_1(\chi)/K_0(\chi)$ (energy density) are calculated as \cite{Eckart}
\begin{eqnarray}
n&=&N^\alpha U_\alpha, \nonumber \\
\Pi^{\left<\alpha\beta\right>}&=&\left(\Delta_\gamma^\alpha \Delta_\delta^\beta-\frac{1}{3} \Delta^{\alpha\beta} \Delta_{\gamma\delta}\right) T^{\gamma\delta}, \nonumber \\
p+\Pi&=&-\frac{1}{3} \Delta_{\alpha\beta} T^{\alpha\beta},\nonumber \\
q^\alpha&=&\Delta_\gamma^\alpha U_\beta T^{\beta\gamma}, \nonumber \\
e&=&\frac{1}{n} U_\alpha T^{\alpha\beta} U_\beta,
\end{eqnarray}
where $\Delta^{\alpha\beta}=\eta^{\alpha\beta}-U^\alpha U^\beta$ is the projector.\\
\textcolor{red}{$\theta$ is directly calculated using $e:=1/\chi+K_1(\chi)/K_0(\chi)$ in Eq. (40), because $e$ is a function of $\theta$ owing to $\chi=1/\theta$. The pertinent descriptions of $\Pi^{\left<\alpha\beta\right>}$, $\Pi$ and $q^\alpha$ are difficult, because we never postulate the phase space $x:=\int m dt$ in the opinion system and are unable to express $\Pi^{\left<\alpha\beta\right>}$, $\Pi$ and $q^\alpha$ with spacetime gradients of $n$, $\bar{m}$ and $\theta$. Here, we understand that $\Pi^{\left<\alpha\beta\right>}$, $\Pi$ and $q^\alpha$ are nonequilibrium moments in the opinion system.}\\
Finally, Eq. (5) is solved using the direct simulation Monte Carlo (DSMC) method \cite {Yano} using $10^5$ sample peoples.\\
\subsection{Characteristics of cooling process via relativistic inelastic-collisions or external force by the second party}
We investigate the characteristics of the cooling process, which is derived from the relativistic inelastic-collision of opinions between two peoples or concentration of opinions of peoples via the external force by the second (or political) party.\\
At first, we consider the cooling process, which is derived from the relativistic inelastic-collision of opinions between two peoples. Consequently, we set $\Lambda=0$, $\Delta=0$ and $B=0$ in Eq. (5). From Eq. (31), we know that the cooling rate depends on $\chi$ and $\mathcal{C}$, when $f=f_{MJ}$. As initial data, we consider two tests. One is Test A-1, in which $f$ is uniformly populated in the range of $0.99 \le \left|m\right|<1$ at $t=0$. The other is Test A-2, in which $f$ is uniformly populated in the range of $0.99\le m <1$ and $-1< m \le -0.9$ at $t=0$. As a result, initial distribution functions in Tests A-1 and A-2 are markedly nonequilibrium. Figure 3 shows temporal evolutions of $\chi$, which are obtained in Tests A-1 and A-2. $\chi$ in Test A-1 is smaller than $\chi$ in Test A-2 at $t=0$, whereas $\chi$ in Tests A-1 and A-2 increase with similar inclinations in the range of $0 \le t \le 6$. The inclination of $\chi$ in Test A-2 decreases at $t \simeq 6$. Provided that $f \sim f_{MJ}$ at $6 \le t$, $\psi_1\left(\chi,\mathcal{C}\right)$ in Test A-2 is smaller than $\psi_1\left(\chi,\mathcal{C}\right)$ in Test A-1. The cooling rate parameter $\psi_1\left(\chi,\mathcal{C}\right)$ increases, as $|\mathcal{C}|$ decreases, when $1<\chi$, as shown in Fig. 1. Provided that $|\mathcal{C}|$ in Test A-1 is smaller than $|\mathcal{C}|$ in Test A-2, $\psi_1\left(\chi,\mathcal{C}\right)$ in Test A-1 is larger than $\psi_1\left(\chi,\mathcal{C}\right)$ in Test A-2, when $f=f_{MJ}$ and $1<\chi$. In later discussion, we, however, find that $\mathcal{C}=U^1K_2\left(\chi\right)/K_1\left(\chi\right)$ is not the temporal constant owing to $f \neq f_{MJ}$, because $T^{01}$ $\left(\neq T_E^{01}\right)$ includes effects of nonequilibrium terms such as $\Pi$, $q^\alpha$, and $\Pi^{\alpha\beta}$ \cite{Eckart}.
\begin{center}
\includegraphics[width=0.5\textwidth]{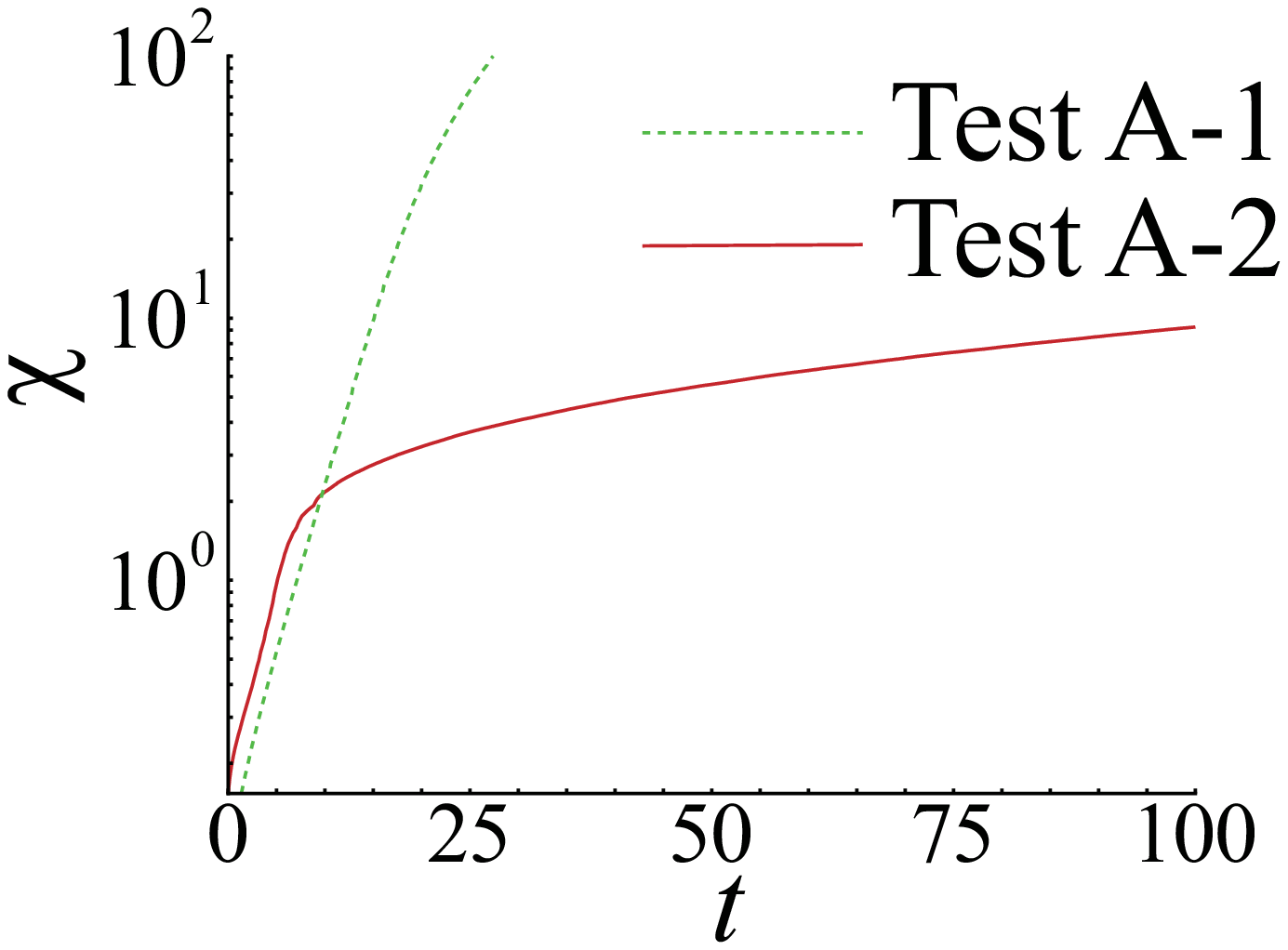}\\
\footnotesize{FIG. 3: Temporal evolutions of $\chi$ in Tests A-1 and A-2.}
\end{center}
Figure 4 shows $f$ and $f_{MJ}$ versus $m$ at $t=17.7$ in Test A-1 (left frame) and $f$ and $f_{MJ}$ versus $m$ at $t=177$ in Test A-2 (right frame). The left frame of Fig. 4 shows that $f$ has higher tails than $f_{MJ}$ at $0.575 \le |m|$. Such higher tails are obtained under the homogeneous cooling state of the granular gas \cite{Brilliantov}. Meanwhile, $f$ has a higher tail at $0.99 \le m <1$ and lower tail at $m \le 0.965$. The initial marked differences between $f$ and $f_{MJ}$ are reduced by relativistic inelastic-collisions in Tests A-1 and A-2.
\begin{center}
\includegraphics[width=0.8\textwidth]{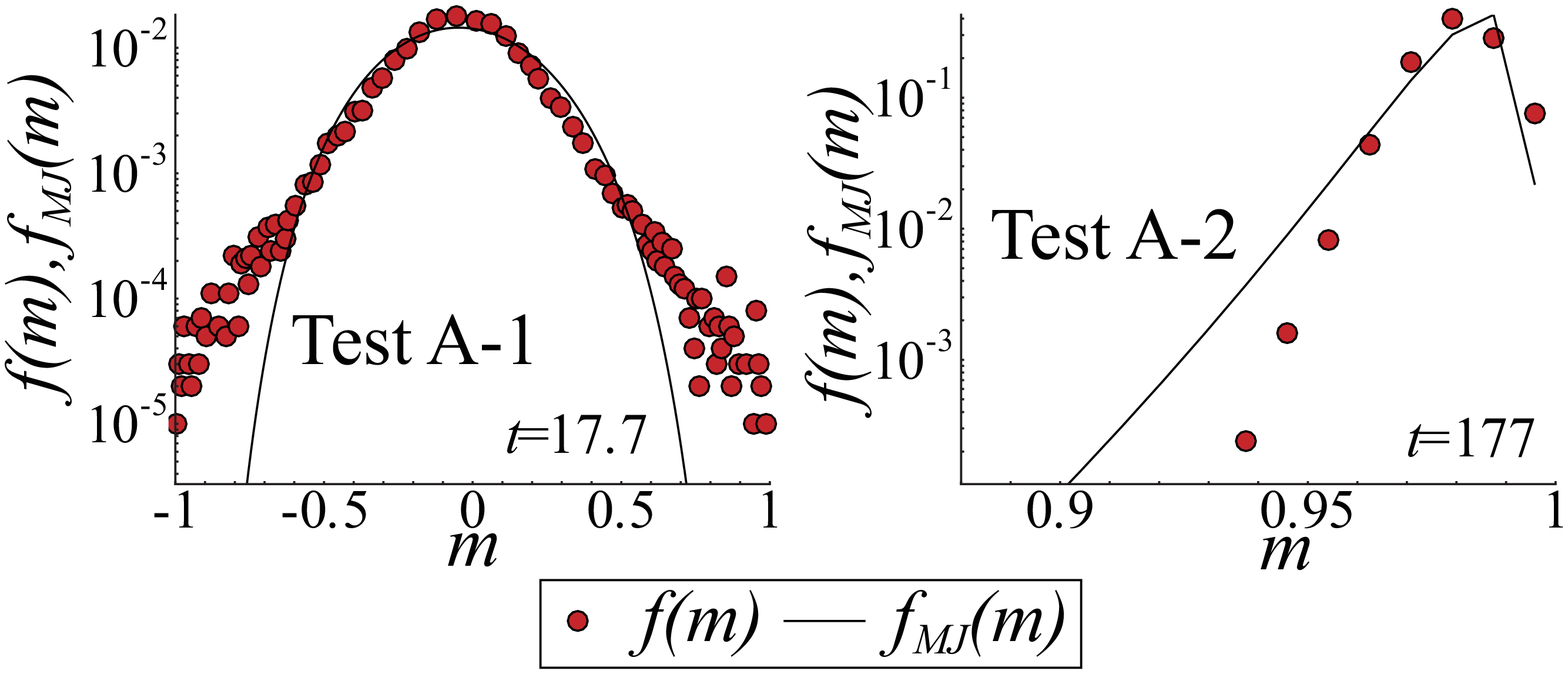}\\
\footnotesize{FIG. 4: $f(m)$ and $f_{MJ}(m)$ versus $m$ at $t=17.7$ in Test A-1 (left frame). $f(m)$ and $f_{MJ}(m)$ versus $m$ at $t=177$ in Test A-2 (right frame).}
\end{center}
Next, we consider the cooling process of $\theta$ via the external force by the second (or political) party by setting $\Lambda=1$ and $\Delta=0$ and $B=0.1$ in Eq. (5). We set $m_p=0$ in Test A-3, where $f$ is uniformly populated in the range of $0.99 \le \left|m\right|<1$ at $t=0$. Additionally, we set $m_p=0.97$ in Test A-4, where $f$ is uniformly populated in the range of $0.99 \le m <1$ and $-1< m \le -0.8$ at $t=0$. Figure 5 shows temporal evolutions of $\chi$, which are obtained in Tests A-3 and A-4. $\chi$ in Tests A-3 and A-4 increase with similar inclinations in the range of $0 \le t \le 10$, whereas $\chi$ in Test A-4 decreases in the range of $10< t \le 20$. Such a decrease of $\chi$ indicates that the external force by the second (or political) party acts as a heating term of $\theta$ owing to the nonequilibrium state of $f$, because the external force by the second (or political) party always acts as a cooling term of $\theta$, when $f=f_{MJ}$. Meanwhile, $\chi$ in Test A-4 increases in the range of $20 < t$ and becomes larger than $\chi$ in Test A-3 at $t \simeq 55$. Such a difference of the cooling process between Tests A-3 and A-4 is caused by the difference of nonequilibrium states between Tests A-3 and A-4. Actually, $f$ in Tests A-3 and A-4 are strongly nonequilibrium at $0 \le t$, as shown in left and right frames of Fig. 6. Consequently, we cannot apply the cooling process in Eq. (39), which postulates $f=f_{MJ}$, to cooling processes in Tests A-3 and A-4. Provided that $f=f_{MJ}$, the larger $|m_p|$ must yield the more rapid cooling of $\theta$. Additionally, initial nonequilibrium states of $f$ never be reduced by the wave propagation term, namely, $p^0 \partial \left(p-P\right)f/\partial p$ in the right hand side of Eq. (5) unlike the relativistic inelastic-collisional term. Meanwhile, an addition of the diffusion term such as $B^\prime(p) p^0 \partial \phi(p) f(p)/\partial p^2$, in which $B^\prime(p)$ and $\phi(p)$ are discussed by Dunkel and Hanggi \cite{Dunkel}, to $B p^0 \partial \left(p-P\right)f(p)/\partial p$ might satisfy H theorem as one dimensional relativistic Fokker Planck equation.
\begin{center}
\includegraphics[width=0.5\textwidth]{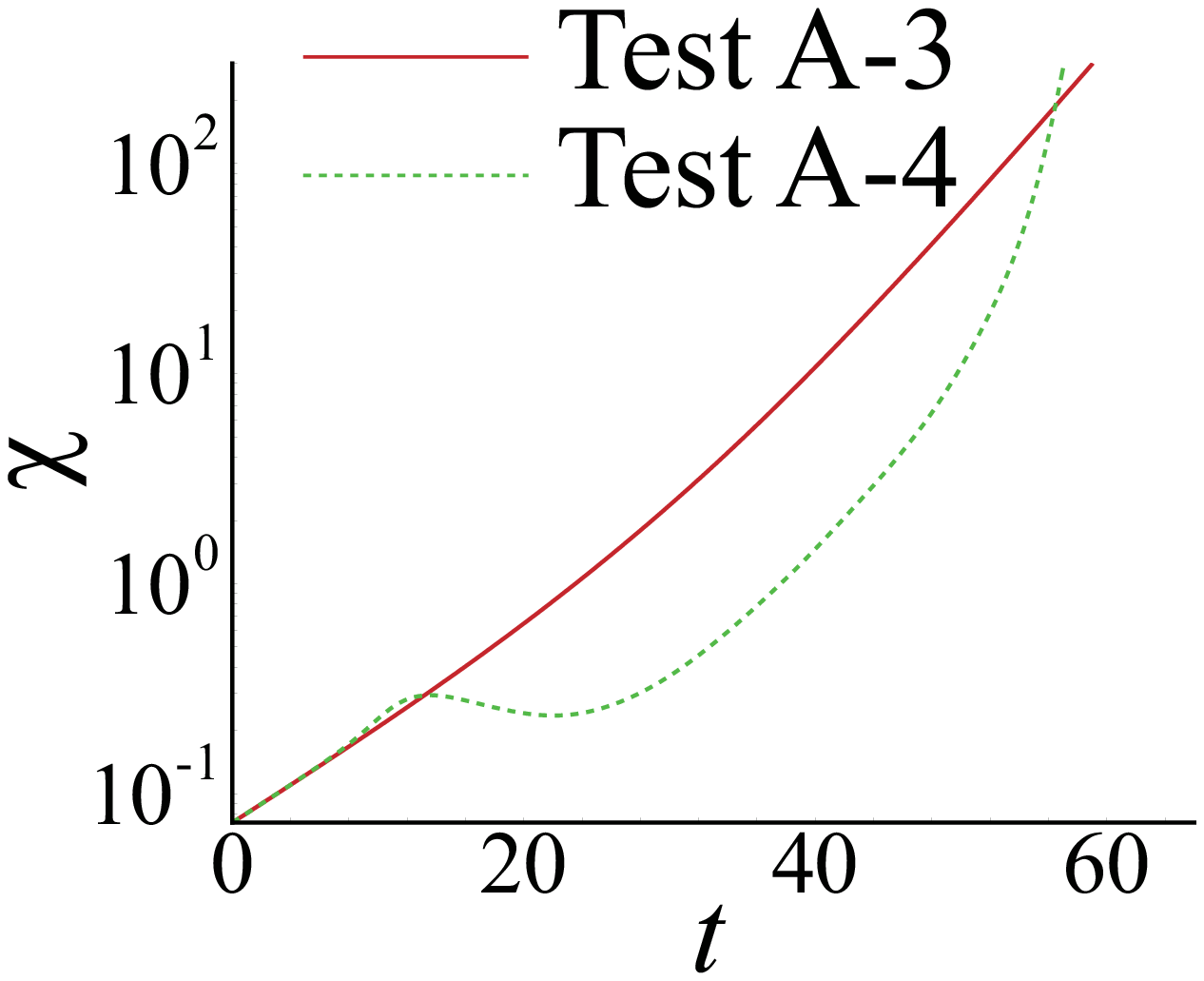}\\
\footnotesize{FIG. 5: Temporal evolutions of $\chi$ in Tests A-3 and A-4.}
\end{center}
Next, we investigate the cooling process of $\theta$, which is analytically obtained in Eq. (31) or (39). Meanwhile, calculations of general solutions of $\chi$ in Eqs. (31) and (39) are markedly difficult. Therefore, we calculate limiting solutions of $\chi$ in Eqs. (31) and (39). Provided that $\mathcal{C}=0$ ($\bar{m}_{t=0}=0$) in Eq. (31), we obtain solutions of $\chi$ under two limiting cases, namely, $\chi \rightarrow 0$ and $\chi \rightarrow \infty$ from Eq. (32) as
\begin{eqnarray}
\chi(t)&=&\chi(0) \exp\left(\frac{A}{8} t\right)~~~~(\chi \rightarrow 0). \\
&=&\left(A\frac{t}{\sqrt{\pi}}+\sqrt{\chi(0)}\right)^2~~~~(\chi \rightarrow \infty).
\end{eqnarray}
Similarly, we obtain solutions of $\chi$ under two limiting cases, namely, $\chi \rightarrow 0$ and $\chi \rightarrow \infty$ from Eq. (39) as
\begin{eqnarray}
\chi(t)&=&\chi(0) \exp\left(B t\right)~~~~(\chi \rightarrow 0). \\
&=&\chi(0) \exp\left(2 B t\right)~~~~(\chi \rightarrow \infty).
\end{eqnarray}
Equations (41) and (42) are plotted together with $\chi$ in Test A-1 in the left frame of Fig. 7, where $A=1$ is used in Eqs. (41) and (42). $\chi_{t=0}$ in Eq. (41) is defined by $\chi_{t=0}$ in Test A-1 and $\chi_{t=0}$ in Eq. (42) is defined by $\chi_{t=15}$ in Test A-1. Eq. (41) gives a good agreement with $\chi$ in Test A-1 in the range of $0 \le t \le 0.5$, whereas Eq. (42) gives a good agreement with $\chi$ in Test A-1 in the range of $15 \le t$. Such good agreements are obtained, because initial nonequilibrium of $f$ is reduced by the relativistic inelastic-collisions, as shown in the left frame of Fig. 4.
\begin{center}
\includegraphics[width=0.8\textwidth]{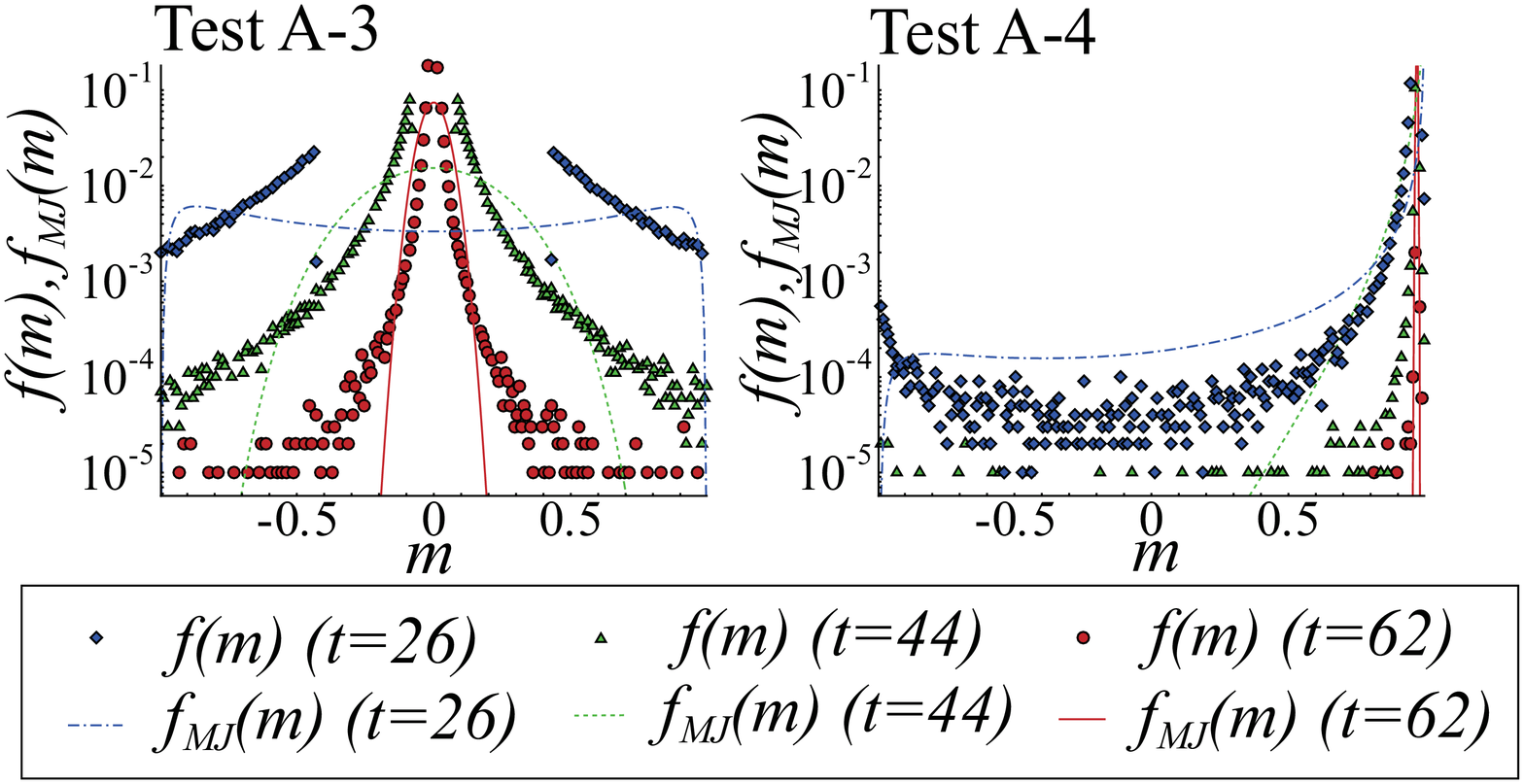}\\
\footnotesize{FIG. 6: Snapshots of temporal evolutions of $f$ and $f_{MJ}$ versus $m$ in Test A-3 (left frame) and Test A-4 (right frame).}
\end{center}
Equations. (43) and (44) are plotted together with $\chi$ in Test A-3 in the right frame of Fig. 7, where $B=0.1$ is used in Eqs. (43) and (44). $\chi_{t=0}$ in Eq. (43) is defined by $\chi_{t=0}$ in Test A-3 and $\chi_{t=0}$ in Eq. (44) is defined by $\chi_{t=40}$ in Test A-3. Eqs. (43) and (44) markedly overestimate the cooling rate in comparison of that in Test A-3, because $f$ in Test A-3 is strongly nonequilibrium at $0 \le t$, as shown in the left frame of Fig. 6.\\
Next, we investigate the temporal evolution of the mean opinion $\bar{m}$ in Tests A-1 - A-4. The left frame of Fig. 8 shows temporal evolutions of $\bar{m}$ in Tests A-1 and A-2, whereas the right frame of Fig. 8 shows temporal evolutions of $\bar{m}$ in Tests A-3 and A-4. $\bar{m}$ in Test A-1 temporally decreases, whereas $\bar{m}$ in Test A-2 temporally increases, as shown in the left frame of Fig. 8.
\begin{center}
\includegraphics[width=0.8\textwidth]{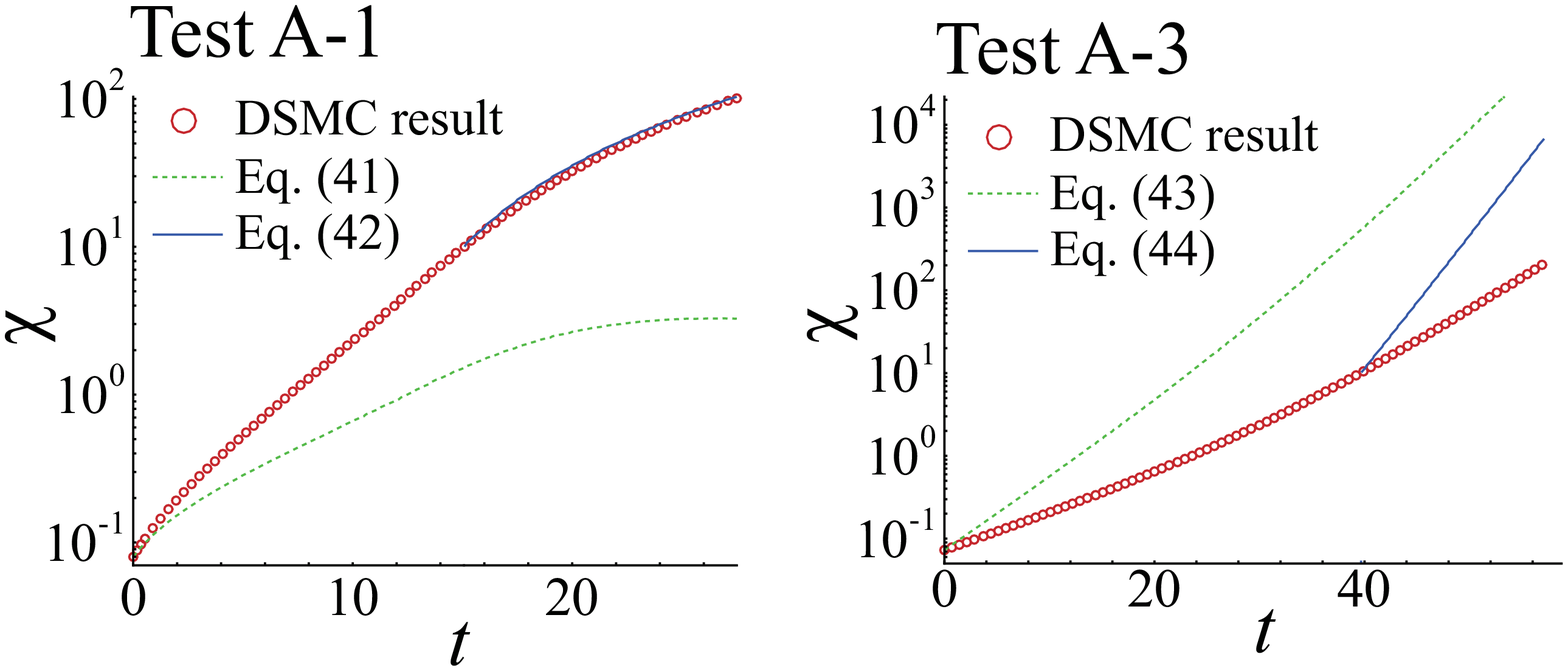}\\
\footnotesize{FIG. 7: Plots of Eqs. (41) and (42) together with temporal evolution of $\chi$ in Test A-1 (left frame). Plots of Eqs. (43) and (44) together with temporal evolution of $\chi$ in Test A-3 (right frame).}
\end{center}
As described in Remark 1, $|\bar{m}|$ increases, as $\chi$ increases. Here, we must remind that $\bar{m}_{t=0}=-\epsilon$ ($0<\epsilon \ll 1$) in Test A-1, which is caused by the thermal fluctuation at $t=0$, yields the temporal evolution of $\bar{m}$ in the range of $\bar{m}_{t=0}<0$. Therefore, the signature of $\bar{m}$ depends on the signature of $\bar{m}_{t=0}$. $\bar{m}$ in Test A-2 approximates to 0.97, as $t$ increases. Such a increase of $\bar{m}$ via the cooling process of $\theta$ indicates that opinions of peoples move toward the decision making state by relativistic inelastic-collisions of opinions between two peoples, unless the initial mean opinion $\bar{m}_{t=0}$ is neutral, namely, $\bar{m}_{t=0}=0$ and $f=f_{MJ}$ for $0 \le t$. In particular, the initial signature of $\bar{m}_{t=0}$, namely, initial bias of the mean opinion to the agreement ($0<\bar{m}_{t=0}$) or disagreement ($\bar{m}_{t=0}<0$) determines the final signature of $\lim_{t \rightarrow \infty} \bar{m}$, namely, final bias of the mean opinion to the agreement ($0<\lim_{t \rightarrow \infty} \bar{m}$) or disagreement ($\lim_{t \rightarrow \infty} \bar{m}<0$).
\begin{center}
\includegraphics[width=1.0\textwidth]{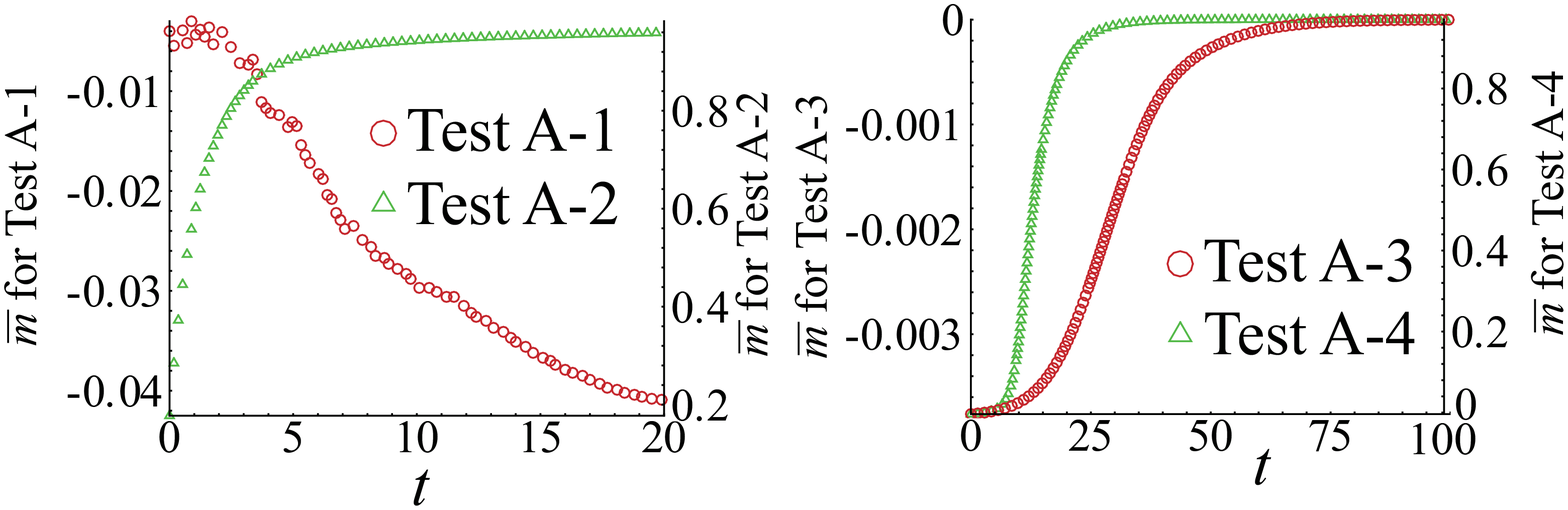}\\
\footnotesize{FIG. 8: Temporal evolutions of $\bar{m}$ in Test A-1 and A-2 (left frame) and temporal evolutions of $\bar{m}$ in Test A-3 and A-4 (right frame).}
\end{center}
The right frame of Fig. 8 indicates that $\bar{m}$ temporally converges to $m_p$ in Tests A-3 and A-4.\\
Finally, we must mention to effects of nonequilibrium of $f$, briefly. Figure 9 shows temporal evolutions of $\Pi$, $q^1$, $\Pi^{\left<11\right>}$ and $U^1K_1(\chi)/K_2(\chi)$, which must be a temporally constant, namely, $\mathcal{C}$, when $f=f_{MJ}$, in Test A-1 (left frame) and Test A-2 (right frame). As shown in the left frame of Fig. 9, $U^1K_1(\chi)/K_2(\chi)$ increases in the range of $0 \le t \le 12$ owing to nonequilibrium effects and approximates to the constant value in the range of $12 <t$. Additionally, the left frame of Fig. 9 indicates that $\Pi$, $q^1$ and $\Pi^{\left<11\right>}$ are significant in the range of $0 \le t \le 12$ and temporally damped to zero in the range of $12<t$ with similar damping rates. As shown in the right frame of Fig. 9, $U^1K_1(\chi)/K_2(\chi)$ increases in the range of $0 \le t \le 2.5$ owing to nonequilibrium effects, decreases in the range of $2.5 < t \le 8.8$, and approximates to the constant value in the range of $8.8 <t$. Additionally, $\Pi$ and $q^1$ are temporally damped to 0 in the range of $6 \le t$ with similar damping rates, whereas $\Pi^{\left<11\right>}$ is temporally damped with a slower damping rate than damping rates of $\Pi$ and $q^1$.\\
Provided that we can assume $f \sim f_{MJ}$ at $6<t$ in Tests A-1 and A-2, $|U^1K_1(\chi)/K_2(\chi)|_{\mbox{\tiny{Test A-1}}} \ll |U^1K_1(\chi)/K_2(\chi)|_{\mbox{\tiny{Test A-2}}}$, which is obtained from the left and right frames of Fig. 9, indicates that $|\mathcal{C}|_{\mbox{\tiny{Test A-1}}} \ll |\mathcal{C}|_{\mbox{\tiny{Test A-2}}}$ might prove $(\psi_1)_{\mbox{\tiny{Test A-2}}}<(\psi_1)_{\mbox{\tiny{Test A-1}}}$ at $6<t$ in Fig. 3 by setting $t=6$ to $t=0$.
\begin{center}
\includegraphics[width=1.0\textwidth]{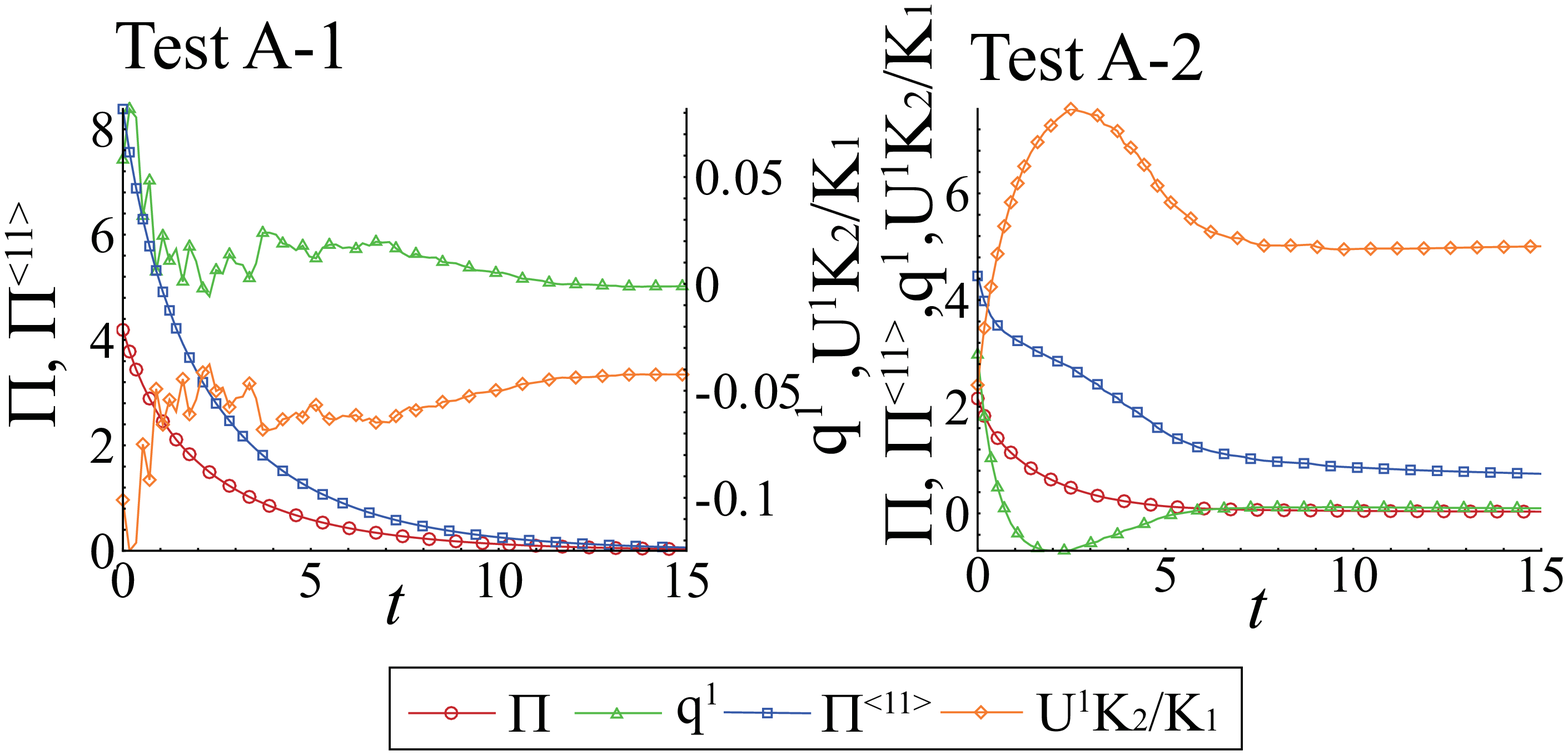}\\
\footnotesize{FIG. 9: Temporal evolutions of $\Pi$, $q^1$, $\Pi^{\left<11\right>}$, and $U^1K_1(\chi)/K_2(\chi)$ in Test A-1 (left frame) and Test A-2 (right frame).}
\end{center} 
\subsection{Characteristics of heating process via self-thinking}
We investigate the characteristics of the heating process via the self-thinking by neglecting effects of the external force by the second (or political) party, namely, $B=0$ and setting $\Lambda=1$ in Eq. (5). Randomly perturbed motion is formulated as $\Delta=\Delta_a \left(2\mathscr{W}-1\right)$, where $\Delta_a$ is the amplitude of randomly perturbed motion, and $0 \le \mathscr{W} \le 1$ is the white noise. Here, we must answer to the question "\textit{Does the global interest eternally increase via the self-thinking, when the binary collision is elastic?"} As initial data, $f$ is uniformly populated in the range of $0 \le |m| <1$ at $t=0$ in Test B-1, whereas $f$ is uniformly populated in ranges of $0.99 \le m <1$ and $-1 < m \le -0.8$ at $t=0$ in Test B-2.
\begin{center}
\includegraphics[width=0.8\textwidth]{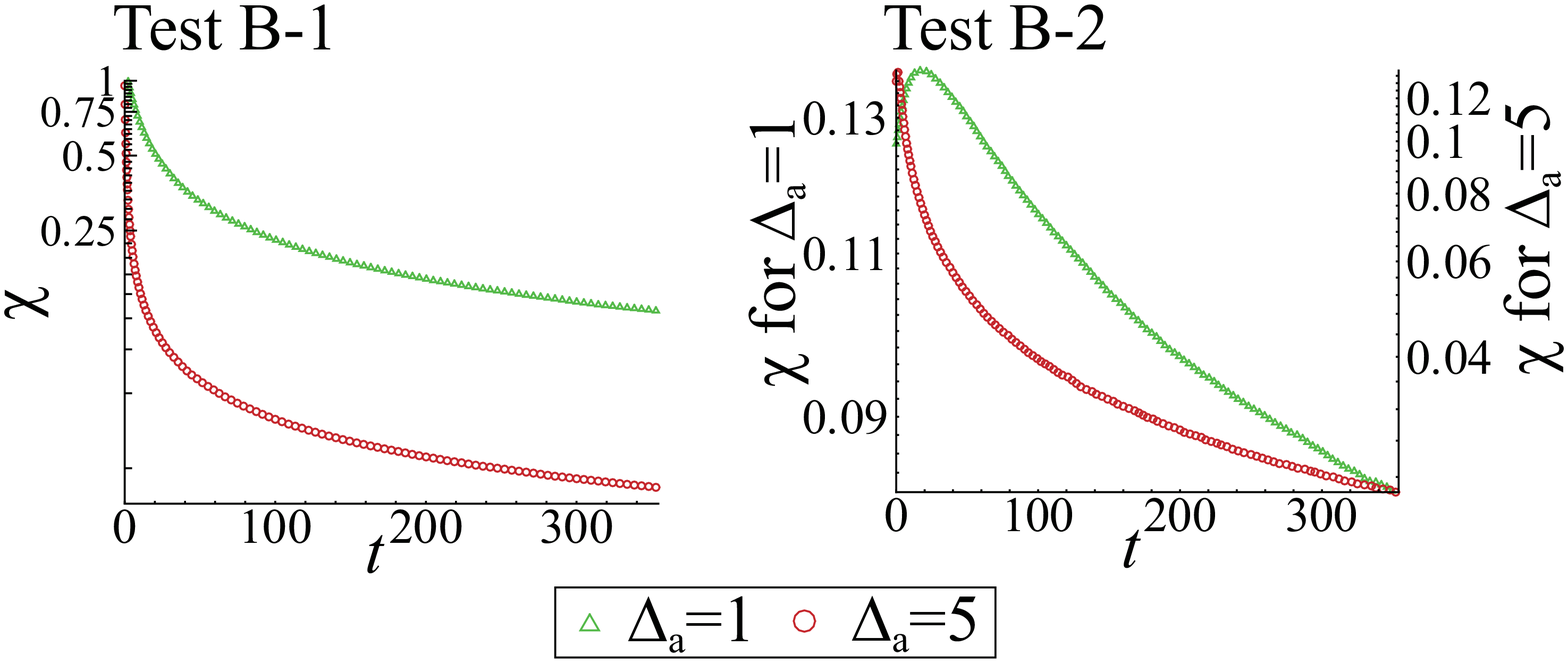}\\
\footnotesize{FIG. 10: Temporal evolutions of $\chi$ in Test B-1 (left frame) and Test B-2 (right frame), when $\Delta_a=1$ and $5$.}
\end{center}
The left frame of Fig. 10 shows temporal evolutions of $\chi$ in Test B-1, when $\Delta_a=1$ and $\Delta_a=5$. $\chi$ ($\theta$) temporally decreases (increases) by randomly perturbed motion. The right frame of Fig. 8 shows temporal evolutions of $\chi$ in Test B-2, when $\Delta_a=1$ and $\Delta_a=5$. $\chi$ ($\theta$), which is obtained using $\Delta_a=5$, temporally increases (decreases) in the range of $0 \le t \le 1$ and decreases in the range of $1<t$ by randomly perturbed motion, whereas $\chi$ ($\theta$), which is obtained using $\Delta_a=1$, temporally increases (decreases) in the range of $0 \le t \le 21$ and decreases (increases) in the range of $21<t$ by randomly perturbed motion. As a result, we confirm that $\theta$ decreases by randomly perturbed motion owing to relativistic effects.
\begin{center}
\includegraphics[width=0.8\textwidth]{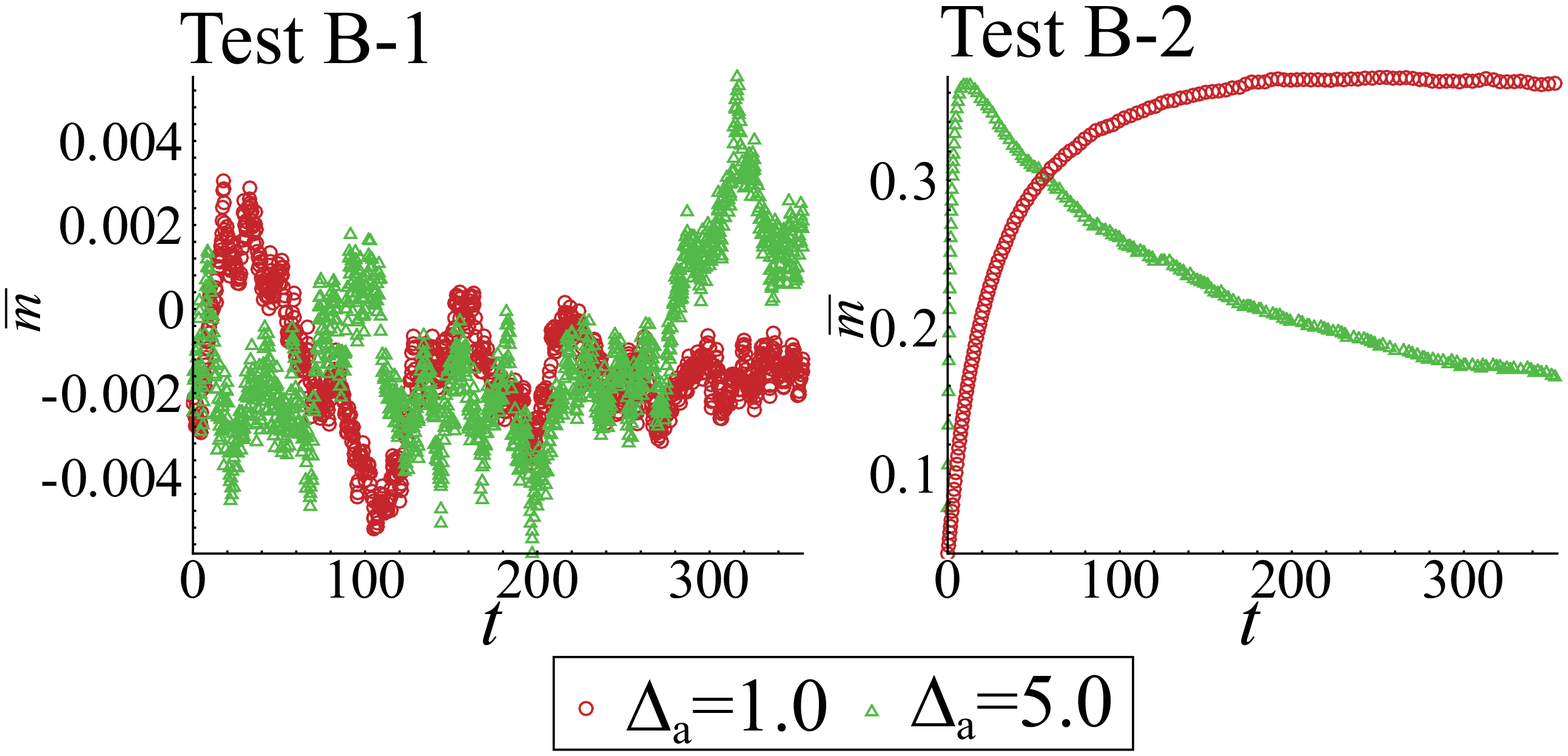}\\
\footnotesize{FIG. 11: Temporal evolutions of $\bar{m}$ in Test B-1 (left frame) and Test B-2 (right frame), when $\Delta_a=1$ and $5$.}
\end{center}
The left frame of Fig. 11 shows temporal evolutions of $\bar{m}$ in Test B-1, when $\Delta_a=1$ and $5$, whereas the right frame of Fig. 9 shows the temporal evolutions of $\bar{m}$ in Test B-2, when $\Delta_a=1$ and $5$. From Remark 1, we know that $|\bar{m}|$ temporally decreases by the heating process, when $f=f_{MJ}$. Meanwhile, $\bar{m}$ in Test B-1 is fluctuating and damped, when $\Delta_a=1$, whereas
$\bar{m}$ in Test B-1 is fluctuating without being damped, when $\Delta_a=5$. The investigation of the autocorrelation of thermal fluctuations of $\bar{m}$ is set to our future study. $\bar{m}$ in Test B-2 increases in the range of $0 \le t \le 10$ and decreases in the range of $10<t$, when $\Delta_a=5$, whereas $\bar{m}$ in Test B-2 increases by $0.37$ and slightly decreases, when $\Delta_a=1$. Such decreases of $\bar{m}$ in Tests B-1 and B-2 are caused by the decrease of $\chi$ from Remark 1, when $f \sim f_{MJ}$. In Test B-2, the range of the increase of $\bar{m}$ is larger than the range of the increase of $\chi$, namely, $0 \le t \le 1$, when $\Delta_a=5$. Meanwhile, we can conclude that $0<d_t |\bar{m}|$ under $d_t \chi <0$ is caused by $f \neq f_{MJ}$, when $\Delta_a=5$. Similarly, $0<d_t |\bar{m}|$ under $d_t \chi <0$ in the range of $21<t$ is caused by $f \neq f_{MJ}$ in Test B-1.
\begin{center}
\includegraphics[width=0.8\textwidth]{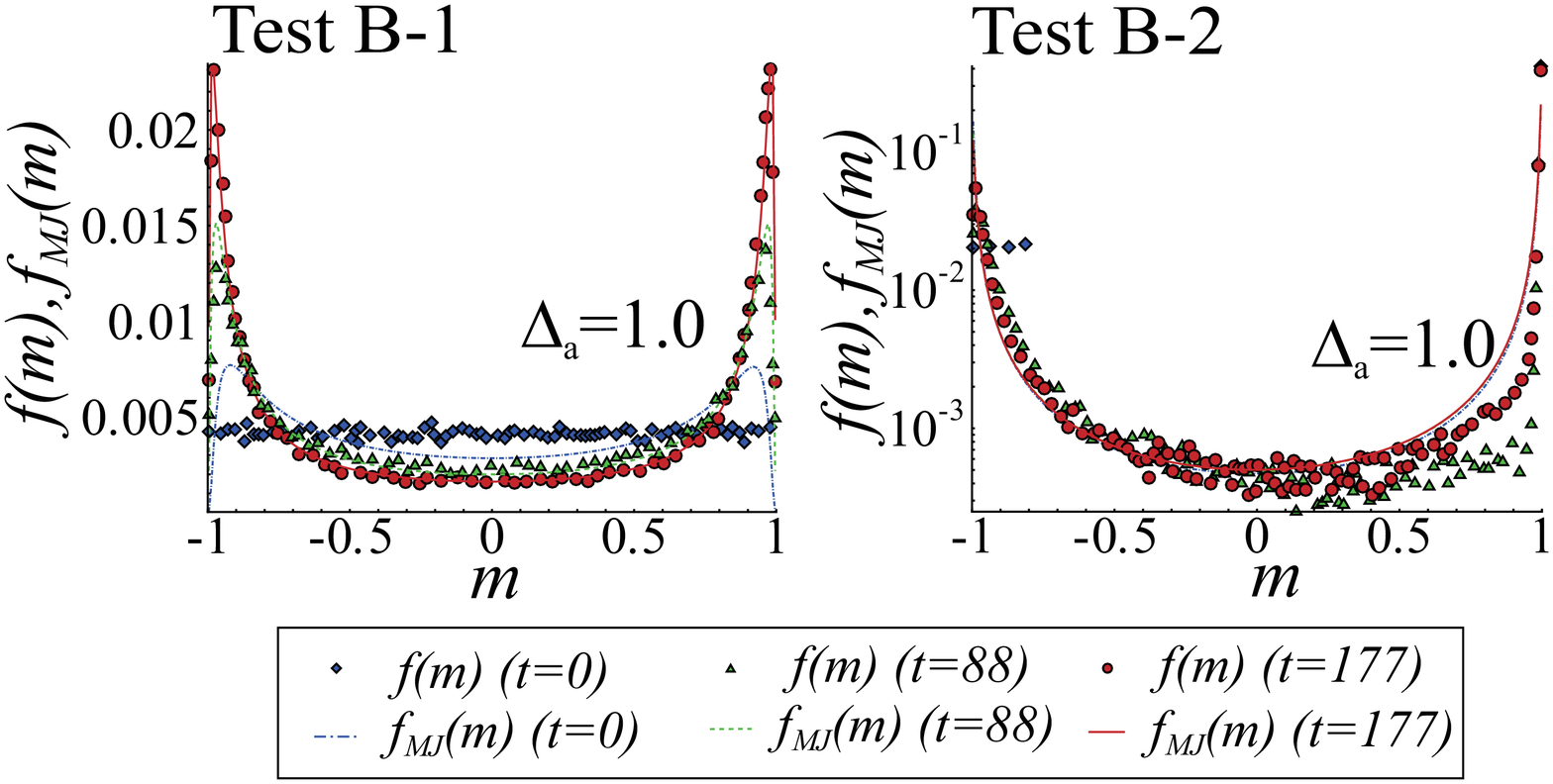}\\
\footnotesize{FIG. 12: Snapshots of temporal evolutions of $f$ and $f_{MJ}$ versus $m$ in Test B-1 (left frame) and Test B-2 (right frame), when $\Delta_a=1$.}
\end{center}
Figure 12 shows snapshots of temporal evolutions of $f$ and $f_{MJ}$ versus $m$ in Test B-1 (left frame) and Test B-2, when $\Delta_a=1$. We can confirm that marked nonequilibrium states of $f$ at $t=0$ in Tests B-1 and B-2 are temporally reduced. The difference between $f$ and $f_{MJ}$ in the range of $0.6 \le m <1$, which is obtained using $\Delta_a=1$ in Test B-2, slightly decreases from $t=88$ to $177$. Finally, opinions of all the peoples approximate to $\pm 1$, namely, complete decision making via the self-thinking under $t \rightarrow \infty$.\\
\subsection{Opinion formation under absence of the second party}
On the basis of discussions in Secs. III-A and B, we investigate the opinion formation via the relativistic inelastic-collision with randomly perturbed motion under the absence of the second (or political) party, when $\Lambda=0$ and $B=0$ in Eq. (5). As initial data, $f$ is uniformly populated in the range of $0.8 \le |m| <1$ at $t=0$ in Test C-1, whereas $f$ is uniformly populated in ranges of $0.99 \le m <1$ and $-1 < m \le -0.8$ at $t=0$ in Test C-2.
\begin{center}
\includegraphics[width=0.7\textwidth]{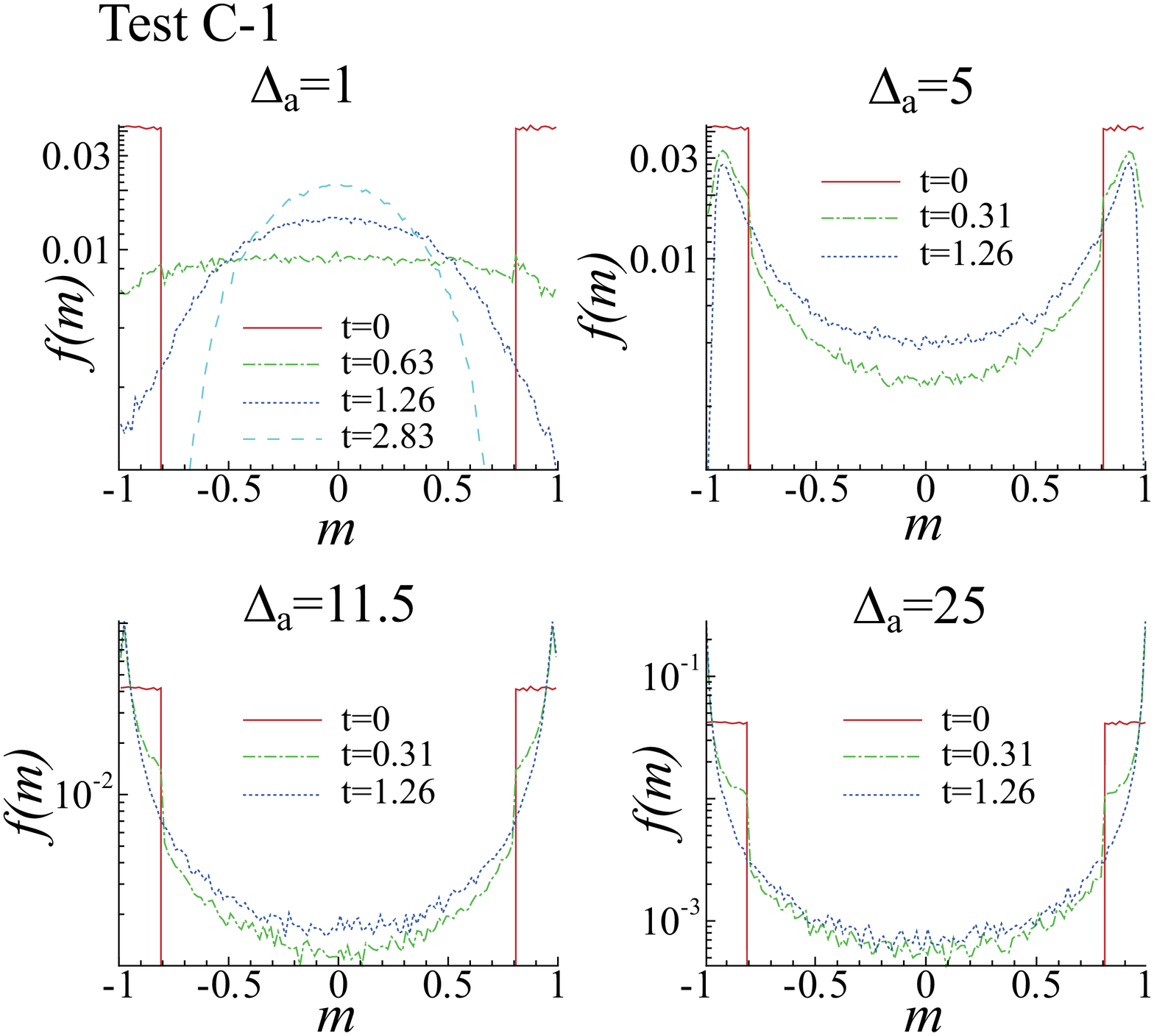}\\
\footnotesize{FIG. 13: Snapshots of temporal evolutions of $f\left(t,m\right)$ versus $m$ in Test C-1, when $\Delta_a=1.0$, $5.0$, $11.5$ and $25$.}
\end{center}
Figure 13 shows snapshots of temporal evolutions of $f\left(t,m\right)$ versus $m$ in Test C-1, which are obtained using $\Delta_a=1.0$, $5.0$, $11.5$ and $25$. $f_{t=1.26}$, which are obtained using $\Delta_a=1.0$, $5.0$, $11.5$ and $25$, are similar to their convergent forms, namely, $f_{t=\infty}$, which are determined by the balance between the cooling rate via the binary exchange of opinions between two peoples and heating rate via the self-thinking. As shown in the top-left frame of Fig. 13, $f_{t=1.26}$ has its peak at $m=0$, when $\Delta_a=1.0$. Consequently, the cooling via the compromise ($\Lambda=0$) in the binary exchange of opinions between two peoples suppresses the heating via the self-thinking, when $\Delta_a=1.0$, because the self-thinking operates as a heating term, when $\bar{m}=0$, as shown in the left frame of Fig. 10. Meanwhile, opinions of peoples move toward the complete decision making state ($m=\pm 1$), as $\Delta_a$ increases, as shown in top-right, bottom-left, and bottom right frames of Fig. 13. As shown in the top-right frame of Fig. 13, $f$ in the high opinion tail, namely, $|m| \sim 1$, temporally decreases, because the cooling via the compromise ($\Lambda=0$) in the binary exchange of opinions between two peoples suppresses the heating via the self-thinking. As shown in bottom-left and bottom-right frames of Fig. 13, $f$ in the high opinion tail, temporally increases.
\begin{center}
\includegraphics[width=0.7\textwidth]{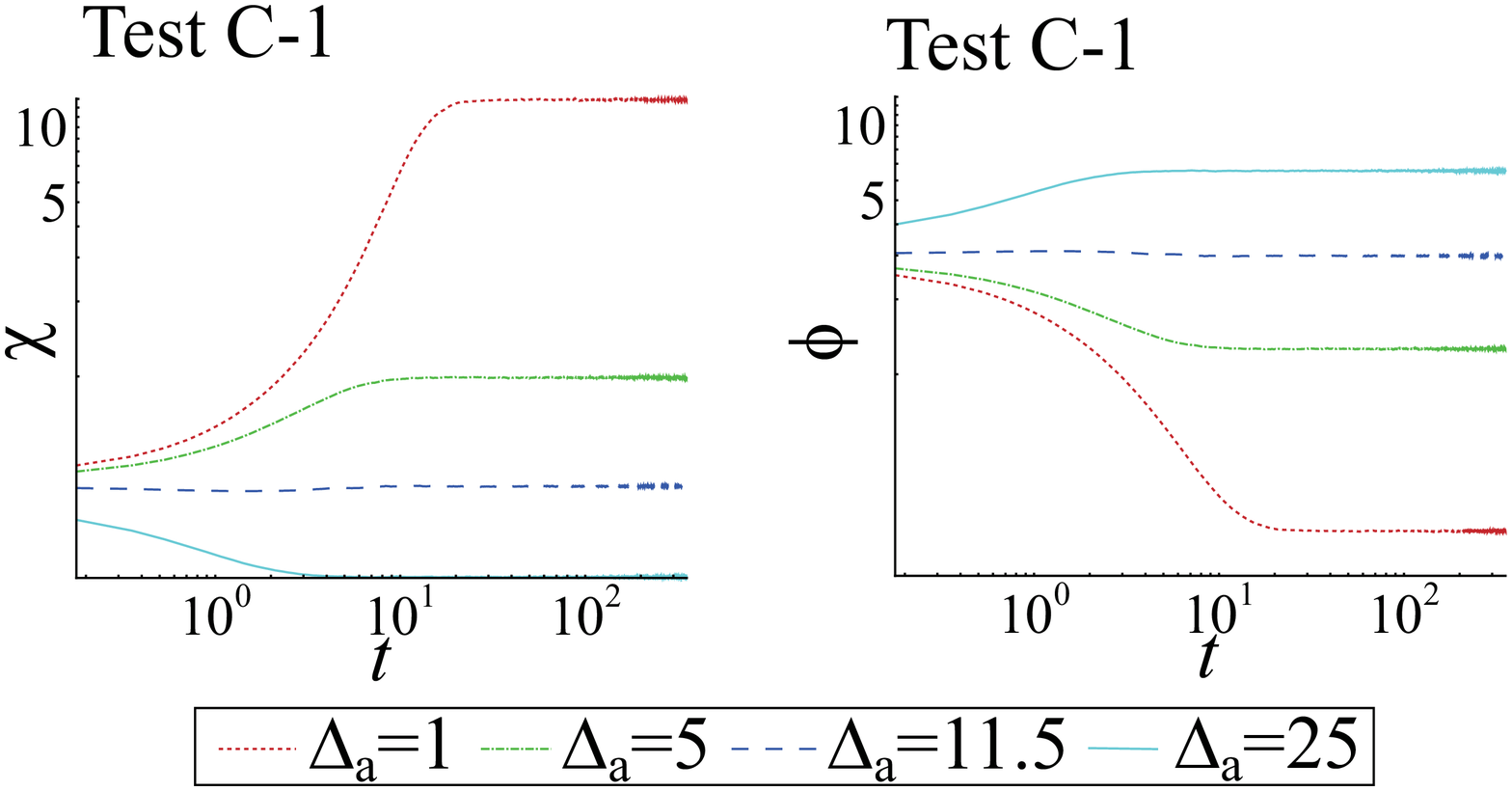}\\
\footnotesize{FIG. 14: Temporal evolutions of $\chi$ (left frame) and $\phi$ (right frame) in Test C-1, when $\Delta_a=1$, $5$, $11.5$ and $25$,.}
\end{center}
\textcolor{red}{To evaluate how opinions approximate to decision making state ($|m|=1$), we introduce the decision making parameter defined by $\phi \left(t\right) := \int_{-\infty}^\infty \left|p\right| f\left(t,p\right) dp$. $\phi$ approximates to $\pm \infty$, when $\theta \rightarrow \infty$ or $\bar{m} \rightarrow \pm 1$.}\\
Figure 14 shows temporal evolutions of $\chi$ (left frame) and $\phi$ (right frame), which are obtained using $\Delta_a=1$, $5$, $11.5$ and $25$. As shown in the left frame of Fig. 14, global interests ($\theta$), which are obtained using $\Delta_a=1$ and $5$, markedly decreases and approximates to their convergent values, whereas $\theta$ obtained using $\Delta_a=11.5$ slightly decreases and $\theta$ obtained using $\Delta_a=25$ markedly increases and approximates to its convergent value. Similarly, decision making parameters ($\phi$), which are obtained using $\Delta_a=1$ and $5$, markedly decrease and approximate to their convergent value, whereas $\phi$ obtained using $\Delta_a=11.5$ slightly decreases, and $\phi$ obtained using $\Delta_a=25$ markedly increases and approximates to its convergent value, as shown in the right frame of Fig. 14. Finally, convergent rates of $\chi$ and $\phi$ increase, as $\Delta_a$ increases. In particular, the increase of the convergent rate of $\chi$ might be described by the cooling rate, which decreases in the range of $\chi<4$, as shown in the right frame of Fig. 1.\\
Next, we investigate the nonequilibrium state of $f_{t=\infty}$ by comparing $f_{t=\infty}$ with the equilibrium distribution function, namely, Maxwell-J$\ddot{\mbox{u}}$ttner function, $\left(f_{MJ}\right)_{t=\infty}$, where we calculate $f$ using the temporal average of $f$, after $f$ approximates to its convergent form. As discussed above, $f$ approximates to its convergent form in accordance with $\Delta_a$, when the cooling rate via the compromise and heating rate via the self-thinking are balanced. We, however, conjecture that $f_{t=\infty}$ never approximates to $\left(f_{MJ}\right)_{t=\infty}$, because the total collisional energy, namely, $E+E_\ast$, is not conserved in the inelastic binary collision with randomly perturbed motion.
\begin{center}
\includegraphics[width=0.7\textwidth]{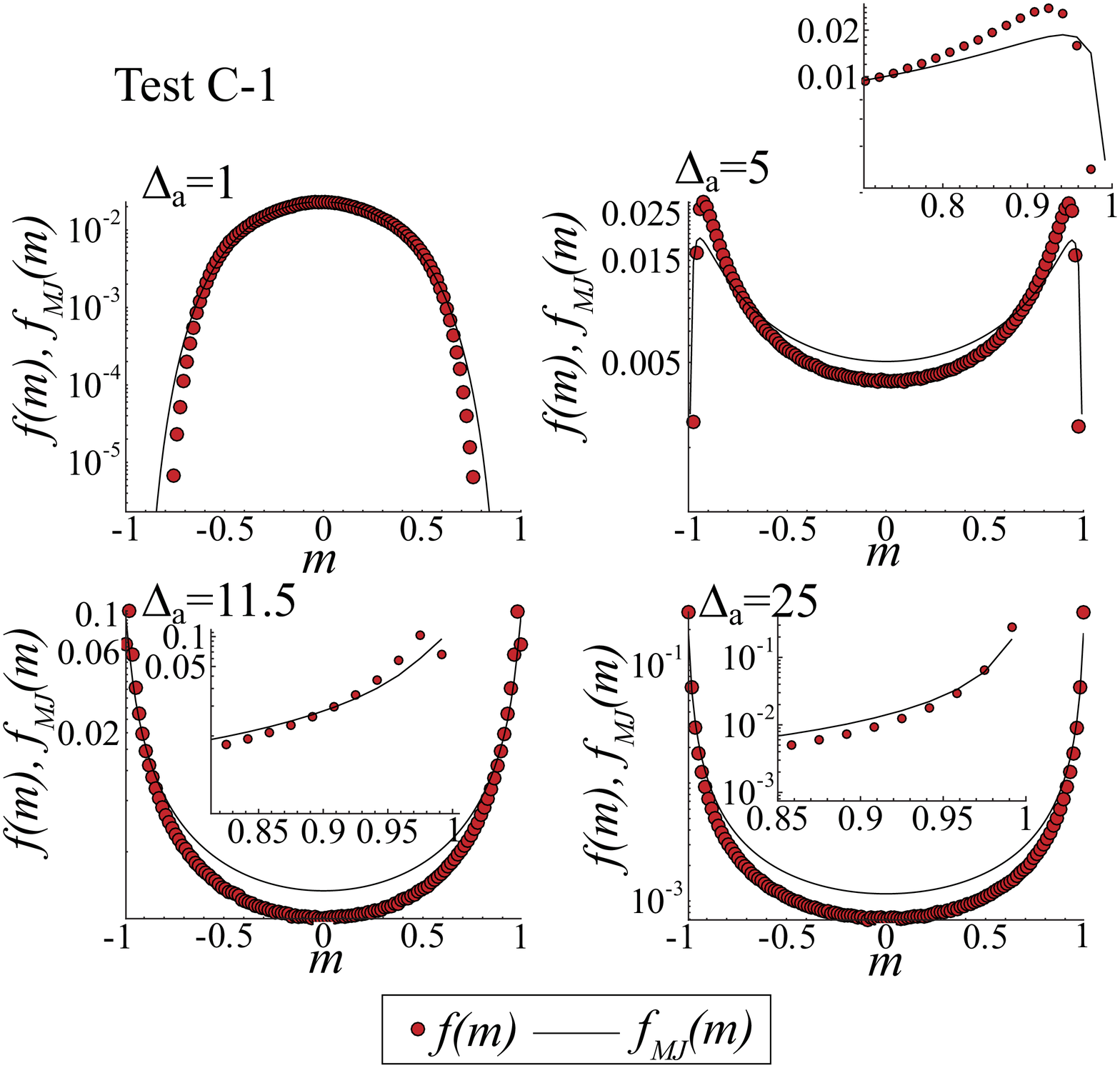}\\
\footnotesize{FIG.15: $f_{t=\infty}$ and $\left(f_{MJ}\right)_{t=\infty}$ versus $m$ in Test C-1, when $\Delta_a=1$ (top-left frame), $\Delta_a=5$ (top-right frame), $\Delta_a=11.5$ (bottom-left frame) and $\Delta_a=25$ (bottom-right frame).}
\end{center}
Figure 15 shows $f_{t=\infty}$ and $\left(f_{MJ}\right)_{t=\infty}$ versus $m$, when $\Delta_a=1$ (top-left frame), $\Delta_a=5$ (top-right frame), $\Delta_a=11.5$ (bottom-left frame) and $\Delta_a=25$ (bottom-right frame). As shown in the top-left frame of Fig. 15, $f_{t=\infty} \le \left(f_{MJ}\right)_{t=\infty}$ in the range of $0 \le |m| \le 0.3$, $\left(f_{MJ}\right)_{t=\infty} \le f_{t=\infty}$ in the range of $0.3 \le |m| \le 0.58$ and $f_{t=\infty} \le \left(f_{MJ}\right)_{t=\infty}$ in the range of $0.58 \le |m| <1$, when $\Delta_a=1$. Consequently, $f_{t=\infty}$ has lower tail than $\left(f_{MJ}\right)_{t=\infty}$ owing to the heating via the self-thinking, whereas $f_{t=\infty}$ has higher tail than $\left(f_{MJ}\right)_{t=\infty}$ owing to the cooling via the relativistic inelastic-collision, when $\bar{m}=0$ and the heating via the self-thinking is neglected, as shown in the left frame of Fig. 4. As shown in the top-right frame of Fig. 15, $f_{t=\infty} \le \left(f_{MJ}\right)_{t=\infty}$ in the range of $0 \le |m| \le 0.72$, $\left(f_{MJ}\right)_{t=\infty} \le f_{t=\infty}$ in the range of $0.72 \le |m| \le 0.95$ and $f_{t=\infty} \le \left(f_{MJ}\right)_{t=\infty}$ in the range of $0.95 \le |m| <1$, when $\Delta_a=5$. As shown in the bottom-left frame of Fig. 15, $f_{t=\infty} \le \left(f_{MJ}\right)_{t=\infty}$ in the range of $0 \le |m| \le 0.9$, $\left(f_{MJ}\right)_{t=\infty} \le f_{t=\infty}$ in the range of $0.9 \le |m| \le 0.98$ and $f_{t=\infty} \le \left(f_{MJ}\right)_{t=\infty}$ in the range of $0.98 \le |m| <1$, when $\Delta_a=11.5$. As shown in the bottom-right frame of Fig. 15, $f_{t=\infty} \le \left(f_{MJ}\right)_{t=\infty}$ in the range of $0 \le |m| \le 0.97$ and $\left(f_{MJ}\right)_{t=\infty} \le f_{t=\infty}$ in the range of $0.97 \le |m| <1$, when $\Delta_a=25$. Finally, we can conclude that $f_{t=\infty}$ is slightly different from the Maxwell-J$\ddot{\mbox{u}}$ttner function, as a result of relativistic inelastic-collisions with randomly perturbed motion.\\
\begin{center}
\includegraphics[width=0.9\textwidth]{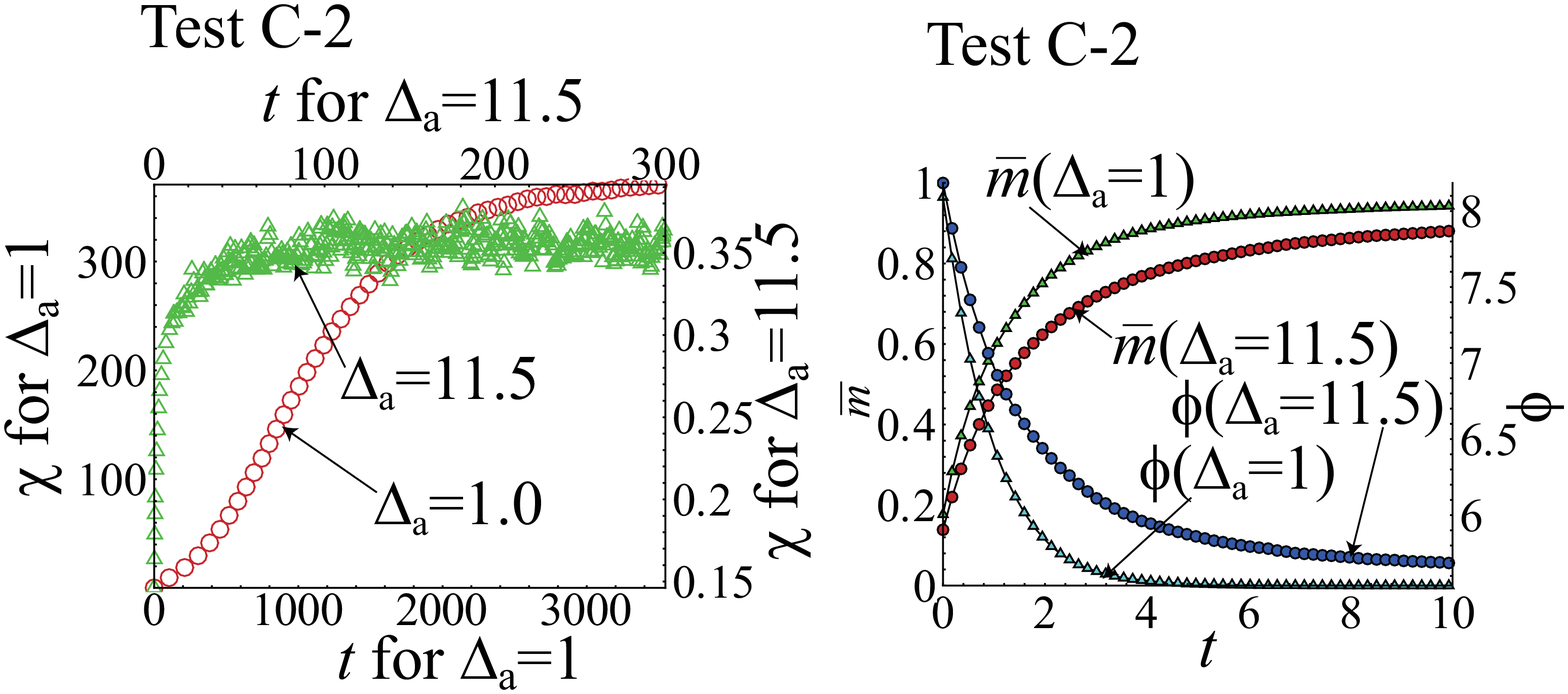}\\
\footnotesize{FIG. 16: Temporal evolutions of $\chi$ in Test C-2, when $\Delta_a=1$ and $11,5$ (left frame). Temporal evolutions of $\bar{m}$ and $\phi$ in Test C-2, when $\Delta_a=1$ and $11.5$ (right frame).}
\end{center}
The left frame of Fig. 16 shows temporal evolutions of $\chi$ in Test C-2, when $\Delta_a=1$ and $11.5$. $\chi$ slightly decreases and approximates to $0.35$, when $\Delta_a=11.5$. On the other hand, $\chi$ markedly decreases and approximates to 360, when $\Delta_a=1$. The right frame of Fig. 16 shows temporal evolutions of $\bar{m}$ and $\phi$ in Test C-2, when $\Delta_a=1$ and $11.5$. $\bar{m}$ increases owing to the decrease of $\theta$ and approximates to constant value, namely, $(\bar{m}_{\Delta_a=1})_{t=\infty}=0.98$ and $(\bar{m}_{\Delta_a=11.5})_{t=\infty}=0.95$, whereas $\phi$ decreases owing to the decrease of $\theta$ and approximates to constant value, namely, $(\phi_{\Delta_a=1})_{t=\infty}=5.54$ and $(\phi_{\Delta_a=11.5})_{t=\infty}=5.58$. We notice that two $\phi$, which are obtained using $\Delta_a=1$ and $11.5$, are similar to each other owing to $(\bar{m}_{\Delta_a=11.5})_{t=\infty}<(\bar{m}_{\Delta_a=1})_{t=\infty}$ despite $(\theta_{\Delta_a=1})_{t=\infty} \ll (\theta_{\Delta_a=11.5})_{t=\infty}$.\\
Figure 17 shows $f_{t=\infty}$ and $\left(f_{MJ}\right)_{t=\infty}$ versus $m$ in Test C-2, when $\Delta_a=11.5$. We obtain $\left(f_{MJ}\right)_{t=\infty} \le f_{t=\infty}$ in the range of $-1< m \le -0.78$, $f_{t=\infty} < \left(f_{MJ}\right)_{t=\infty}$ in the range of $-0.78< m < 0.83$, $\left(f_{MJ}\right)_{t=\infty} \le f_{t=\infty}$ in the range of $0.83 \le m \le 0.988$, and $f_{t=\infty} < \left(f_{MJ}\right)_{t=\infty}$ in the range of $0.988<m<1$. In other words, number of peoples, whose opinions are similar to negative decision making, namely, $m=-1$, is larger than that under the thermally equilibrium state, whereas number of peoples, whose opinions are similar to neutral state, namely, $\left|m\right| \sim 0$ is smaller than that under the thermally equilibrium state, and the number of peoples, whose opinions are quite similar to positive decision making state, namely, $m=1$, is smaller than that under the thermally equilibrium state.
\begin{center}
\includegraphics[width=0.5\textwidth]{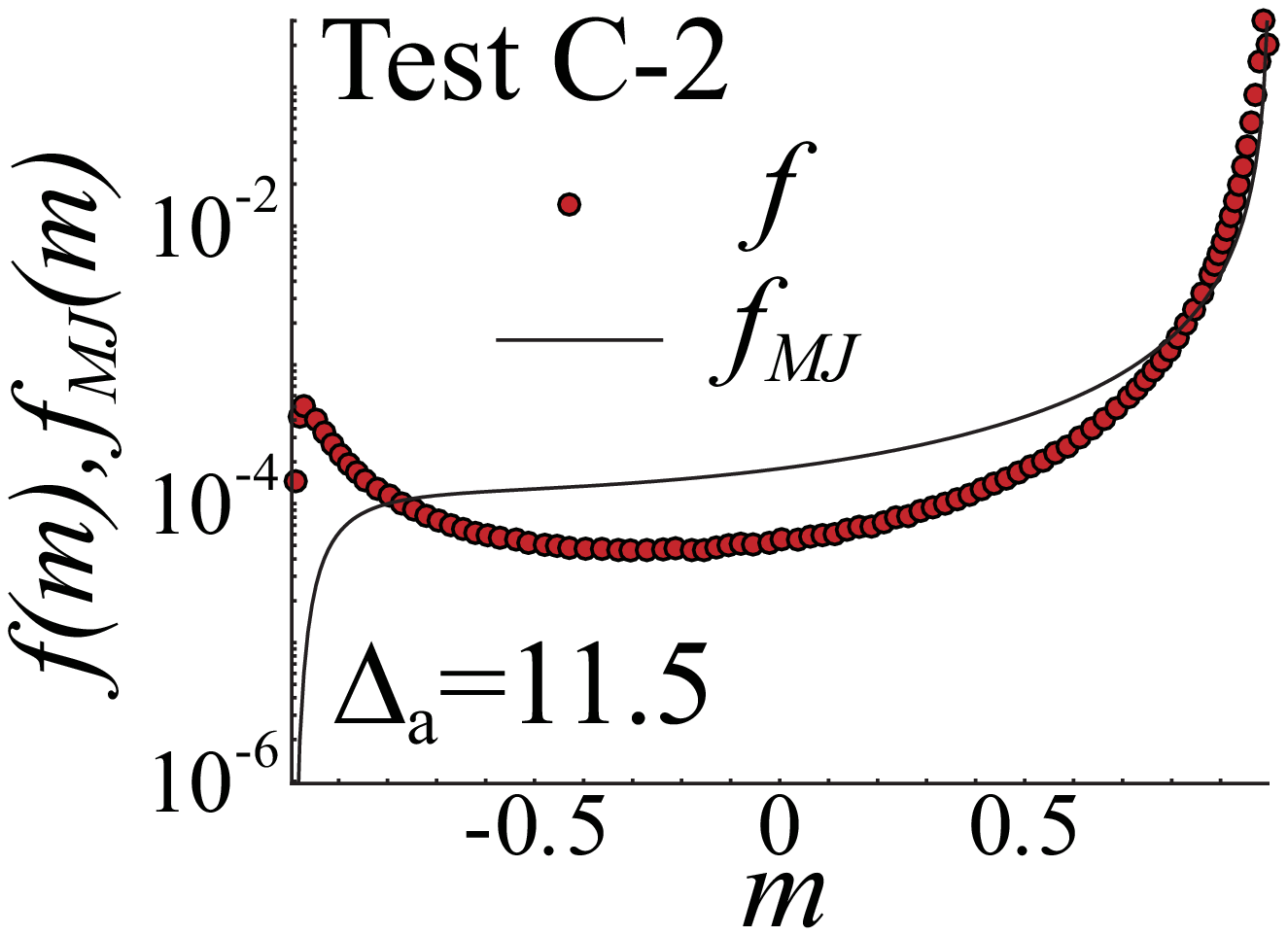}\\
\footnotesize{FIG. 17: $f_{t=\infty}$ and $(f_{MJ})_{t=\infty}$ versus $m$ in Test C-2, when $\Delta_a=11.5$.}
\end{center}
\subsection{Opinion formation under external force by the second party}
Next, we investigate effects of the second (or political) party on the opinion formation using $\Lambda=0$ in Eq. (5). As initial data, $f$ is uniformly populated in the range of $0.8 \le |m| <1$ at $t=0$. We consider six cases, namely, $m_p=0$, $0.5$ and $0.8$, when $\left(\Delta_a,B\right)=\left(1,0.1\right)$, and $m_p=0$, $0.5$ and $0.8$, when $\left(\Delta_a,B\right)=\left(11.5,1\right)$.
\begin{center}
\includegraphics[width=0.8\textwidth]{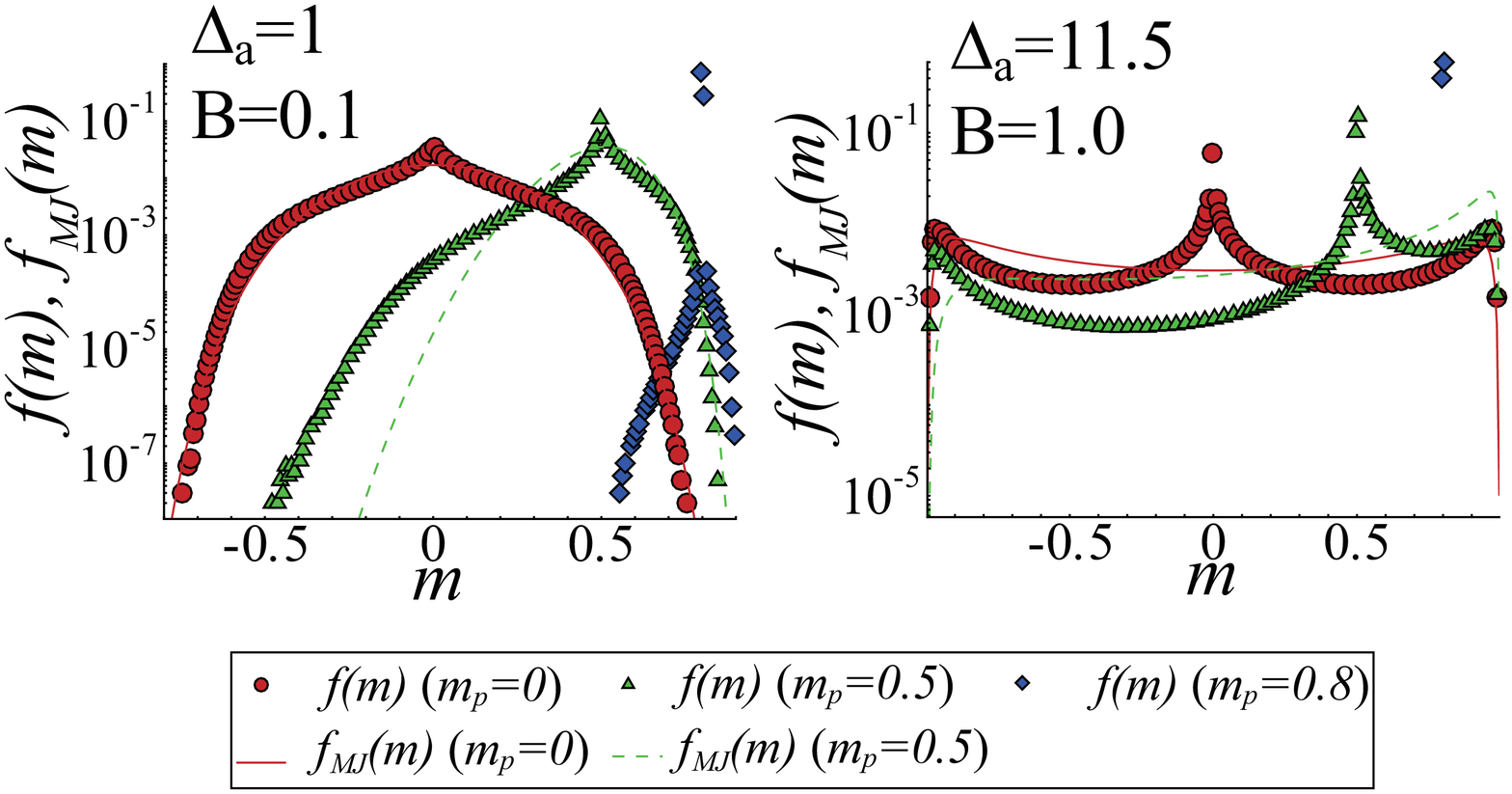}\\
\footnotesize{FIG. 18: $f_{t=\infty}$ versus $m$, for $m_p=0$, $0.5$ and $0.8$, and $\left(f_{MJ}\right)_{t=\infty}$ versus $m$ for $m_p=0$ and $m_p=0.5$, when $\left(\Delta_a,B\right)=(1,0.1)$, in left frame, and $f_{t=\infty}$ versus $m$, for $m_p=0$, $0.5$ and $0.8$, and $\left(f_{MJ}\right)_{t=\infty}$ versus $m$ for $m_p=0$ and $0.5$, when $\left(\Delta_a,B\right)=(11.5,1)$, in right frame.}
\end{center}
Figure 18 shows the convergent form of the distribution function, namely, $f_{t=\infty}$, for $m_p=0$, $0.5$ and $0.8$, and $\left(f_{MJ}\right)_{t=\infty}$ for $m_p=0$ and $m_p=0.5$, when $\left(\Delta_a,B\right)=(1,0.1)$, in its left frame, and $f_{t=\infty}$, for $m_p=0$, $0.5$ and $0.8$, and $\left(f_{MJ}\right)_{t=\infty}$ for $m_p=0$ and $0.5$, when $\left(\Delta_a,B\right)=(11.5,1)$, in its right frame. $\left(f_{MJ}\right)_{t=\infty}$ for $m_p=0.8$ when $\left(\Delta_a,B\right)=\left(1,0.1\right)$ and $\left(\Delta_a,B\right)=\left(11.5,1\right)$, are not shown in Fig. 18, because the modified function of the second kind, which defines $\left(f_{MJ}\right)_{t=\infty}$, cannot be calculated using $\chi$, which is larger than 500, owing to the unestablished algorithm to solve the modified Bessel function of the second kind with $500<\chi$. As shown in left and right frames of Fig. 18, $m$ moves toward $m_p$ owing to the second (or political) party, which is expressed by Vlasov term in Eq. (5). The maximum value of $f_{t=\infty}$ increases, as $m_p$ increases, as shown in the left and right frames of Fig. 18. Consequently, we obtain $f_{t=\infty} \ll 1$ in the range of $m \le -0.5$, when $\left(\Delta_a,B\right)=(1.0,0,1)$ and $m_p=0.5$, and $f_{t=\infty} \ll 1$ in the range of $\chi \le 0.5$, when $\left(\Delta_a,B\right)=\left(1.0,0,1\right)$ and $m_p=0.8$. On the other hand, $f_{t=\infty}$ for $m_p=0$ and $0.5$ have finite values at $|m| \sim 1$ and $f_{t=\infty}$ for $m_p=0.8$ has the markedly sharp peak at $m=m_p$, when $\left(\Delta_a,B\right)=\left(11.5,1\right)$. The sharp $f(m)_{t=\infty}$ and $f\left(0\right)_{m_p=0,t=\infty}<f\left(0.5\right)_{m_p=0.5,t=\infty}<f\left(0.8\right)_{m_p=0.8,t=\infty}$ in left and right frames of Fig. 18 might be described by the fact that the increase of the cooling rate in accordance with the increase of $|P|$ accelerates the decrease of $\theta$, as shown in Figs. 2 and 5, whereas the increase of $\bar{m}(=m_p)$ yields the decrease of the cooling rate of $\theta$ via relativistic inelastic-collisions, as shown in Fig. 3, and the heating rate does not change markedly, as $\bar{m}$ increases, by comparing the left frame of Fig. 10 with the right frame of Fig. 10.\\
From above results, the strong opinion of the second (or political) party concentrates opinions of peoples to the opinion of the second (or political) party, whereas the neutral opinion of the second (or political) party allows opinions, which are different from that of the second (or political) party, when the heating via the self-thinking is large, adequately.
\begin{center}
\includegraphics[width=0.75\textwidth]{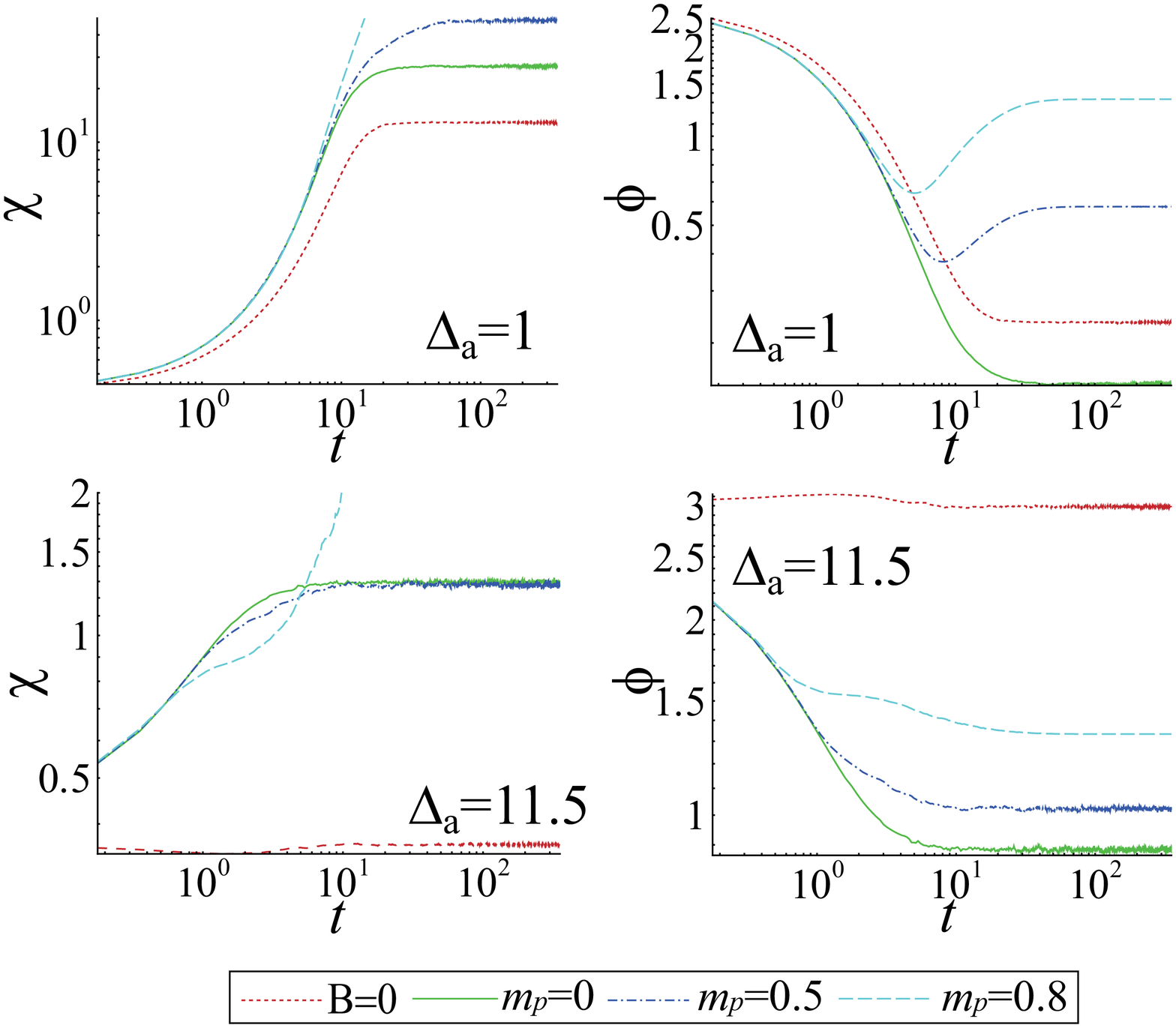}\\
\footnotesize{FIG. 19: Temporal evolutions of $\chi$ (top-left frame) and $\phi$ (top-right frame) for $m_p=0$, $0.5$ and $0.8$, when $(\Delta_a,B)=(1,0.1)$ together with temporal evolutions of $\chi$ and $\psi$, when $(\Delta_a,B)=(1,0)$. Temporal evolutions of $\chi$ (bottom-left frame) and $\psi$ (bottom-right frame) for $m_p=0$, $0.5$ and $0.8$, when $(\Delta_a,B)=(11.5,1)$ together with temporal evolutions of $\chi$ and $\psi$, when $(\Delta_a,B)=(11.5,0)$.}
\end{center}
Figure 19 shows temporal evolutions of $\chi$ (top-left frame) and $\phi$ (top-right frame) for $m_p=0$, $0.5$, and $0.8$, when $\left(\Delta_a,B\right)=\left(1.0,0,1\right)$, together with temporal evolutions of $\chi$ and $\phi$, when $\left(\Delta_a,B\right)=\left(1.0,0\right)$, and temporal evolutions of $\chi$ (bottom-left frame) and $\phi$ (bottom-right frame) for $m_p=0$, $0.5$ and $0.8$, when $\left(\Delta_a,B\right)=\left(11.5,1\right)$, together with temporal evolutions of $\chi$ and $\phi$, when $\left(\Delta_a,B\right)=\left(11.5,0\right)$.\\
As shown in the top-left frame of Fig. 19, the convergent value of $\chi$ ($\theta$) increases (decreases), as $m_p$ increases. Temporal evolutions of $\chi$, which are obtained using $m_p=0,$ $0.5$ and $0.8$, are similar to each other in the range of $0 \le t <5$, when $\left(\Delta_a,B\right)=(1,0.1)$. Similarly, temporal evolutions of $\chi$, which are obtained using $m_p=0,$ $0.5$ and $0.8$, are similar to each other in the range of $0 \le t <0.5$, when $\left(\Delta_a,B\right)=(11.5,1)$. $\chi$, which is obtained using $m_p=0.8$, is smaller $\chi$, which are obtained using $m_p=0$ and $m_p=0.5$, in the range of $0.5 \le t \le 5$, when $\left(\Delta_a,B\right)=(11.5,1)$. Such a tendency is similar to temporal evolutions of $\chi$ in Tests A-3 and A-4, as shown in Fig. 5. Meanwhile, $\chi_{t=\infty}$ ($\theta_{t=\infty}$), which are obtained using $m_p=0$ and $0.5$, when $\left(\Delta_a,B\right)=(11.5,1)$, are markedly higher (lower) than that obtained using $\left(\Delta_a,B\right)=(11.5,0)$. Additionally, $\chi_{t=\infty}$, which is obtained using $m_p=0$, when $\left(\Delta_a,B\right)=(11.5,1)$, is similar to that obtained using $m_p=0.5$, when $\left(\Delta_a,B\right)=(11.5,1)$, whereas $\chi_{t=\infty}$ ($\theta_{t=\infty}$), which is obtained using $m_p=0.8$, when $\left(\Delta_a,B\right)=(11.5,1)$, is markedly higher (lower) than those obtained using $m_p=0$ and $0.5$, when $\left(\Delta_a,B\right)=(11.5,1)$. The relation, $(\theta_{m_p=0.8})_{t=\infty} \ll (\theta_{m_p=0})_{t=\infty} \simeq (\theta_{m_p=0.5})_{t=\infty}$, when $\left(\Delta_a,B\right)=(11.5,1)$, implies that the drastic transition of $\theta_{t=\infty}$ occurs at the critical value of $m_p$, when $\theta_{t=\infty}$ is high enough such as $\chi_{t=\infty} \sim 1$, whereas $\theta_{t=\infty}$ decreases gradually, as $m_p$ increases, when $\theta_{t=\infty}$ is low such as $10<\chi_{t=\infty}$, as shown in the case of $\left(\Delta_a,B\right)=(1,0.1)$. Here, we remind that the increase of $\psi_2$ at $\chi \sim 1$ emerges, as $P$ increases, as shown in Fig. 2. The correlation between the drastic decrease of $\theta$, when $\left(\Delta_a,B\right)=(11.5,1)$, and $\psi_2$ is, however, not clarified, because we did not consider effects of nonequilibrium terms such as $\Pi$, $\Pi^{<\alpha\beta>}$ and $q^\alpha$ on the cooling process of $\theta$.\\
The top-right frame of Fig. 19 shows that the convergent value of the decision making parameter ($\phi_{t=\infty}$), which is obtained using $m_p=0$, when $\left(\Delta_a,B\right)=(1,0.1)$, is smaller than that obtained when $\left(\Delta_a,B\right)=(1,0)$, whereas $\phi_{t=\infty}$ increases, as $m_p$ increases, when $\left(\Delta_a,B\right)=(1,0.1)$. As a result, we obtain the relation, $(\phi_{m_p=0})_{t=\infty}<(\phi_{B=0})_{t=\infty}<(\phi_{m_p=0.5})_{t=\infty}<(\phi_{m_p=0.8})_{t=\infty}$, when $\Delta_a=1$ and $B=0.1$ for $m_p=0$, $0.5$ and $0.8$. The stronger opinion of the second (or political) party yields the larger $\phi_{t=\infty}$, when $\left(\Delta_a,B\right)=(1,0)$. Such a tendency is obtained, when $\left(\Delta_a,B\right)=\left(11.5,1\right)$, whereas we obtain the relation, $(\phi_{m_p=0})_{t=\infty}<(\phi_{m_p=0.5})_{t=\infty}<(\phi_{m_p=0.8})_{t=\infty}<(\phi_{B=0})_{t=\infty}$, when $\Delta_a=11.5$ and $B=1$ for $m_p=0$, $0.5$ and $0.8$.
\section{Concluding remarks}
We investigated the opinion formation of peoples under the second (or political) party with upper and lower bounds to the strength of the opinion using the relativistic inelastic-Boltzmann-Vlasov equation. The temperature of the opinion was regarded as the global interest in the single issue. The decrease of the global interest via the relativistic inelastic-collision yields the increase of the absolute value of the strength of the mean opinion under the thermally equilibrium state. Additionally, the cooling rate of the global interest depends on the initial mean opinion and initial global interest under the thermally equilibrium state, whereas the cooling rate of the global interest via the external force by the second (or political) party depends on the absolute value of the strength of opinion of the second (or political) party under the thermally equilibrium state. Numerical results indicated that effects of nonequilibrium terms on the cooling process were significant under the strongly thermally nonequilibrium state. For example, the cooling process via the external force by the second (or political) party turns into the heating process under the thermally nonequilibrium state, and the heating process via the self-thinking turns into the cooling process under the thermally nonequilibrium state. Provided that we restrict ourselves to the inelastic collisions, the mean opinion moved toward the decision making state, whose signature (positive: agreement, negative: disagreement) coincides with the signature of the initial mean opinion, owing to the inelastic collision. Therefore, the initial small deviation of the mean opinion from zero (neutral state) determines the final state of the opinion. The convergent form of the distribution function was determined by the balance between the cooling rate via the relativistic inelastic-collision and the external force by the second (or political) party, and heating rate via the self-thinking. In particular, the markedly strong opinion of the second (or political) party excluded individuals with opinions, which are different from the opinion of the second (or political) party. The decision making parameter becomes larger by the stronger opinion of the second (or political) party, even when the global interest becomes lower by the stronger opinion of the second (or political) party. \textcolor{red}{As a result, the second (or political party) with the strong opinion, namely, $|m_p| \sim 1$, which decreases the global interests of peoples and lead the mean opinion to the decision making state, is often called as the authoritarian party.}
\begin{acknowledgments}
Authors gratefully acknowledges Dr. Serge
 Galam (CREA and CNRS, Paris) for teaching us his numerous works in sociophysics \cite{Galam} and Professor Gilberto Medeiros Kremer (Dept. of Phys, Univ. Federal do Parana) for teaching us his previous study on (1+1) dimensional relativistic kinetic theory \cite{Kremer}.
\end{acknowledgments}

\begin{appendix}
\section{Definitions of equilibrium moments}
In this appendix, some equilibrium moments are defined.\\
At first, the zeroth order moment is defined as
\begin{eqnarray}
Z=\int_{-\infty}^{\infty} \exp(-\chi p^\alpha U_\alpha) \frac{dp}{p^0}.
\end{eqnarray}
In Lorentz rest frame, we obtain
\begin{eqnarray}
Z=\int_{-\infty}^\infty \exp\left(-\chi p^0 U_0\right) \frac{dp}{p^0}=2 K_0(\chi).
\end{eqnarray}
$Z^{\alpha\beta\gamma\delta...}=\int_{-\infty}^\infty p^\alpha p^\beta p^\gamma p^\delta ... \exp(-\chi p^\alpha U_\alpha) \frac{dp}{p^0}$ is obtained by the successive differentiation of $Z$ using $-\chi U_\alpha$ as
\begin{eqnarray}
&&Z^\alpha=2 K_1(\chi)U^\alpha,\\
&&Z^{\alpha\beta}=2 K_2(\chi)U^\alpha U^\beta-2\eta^{\alpha\beta} \frac{K_1(\chi)}{\chi},\\
&&Z^{\alpha\beta\gamma}=2 K_3(\chi)U^\alpha U^\beta U^\gamma-2\left(\eta^{\alpha\beta}U^\gamma+\eta^{\alpha\gamma}U^\beta+\eta^{\beta\gamma}U^\alpha \right)\frac{K_2(\chi)}{\chi},\\
&&Z^{\alpha\beta\gamma\delta}=2 K_4(\chi)U^{\alpha} U^\beta U^\gamma U^\delta \nonumber \\
&&-\frac{2 K_3(\chi)}{\chi}\left(\eta^{\alpha\beta} U^\gamma U^\delta+\eta^{\alpha\gamma} U^\beta U^\delta+\eta^{\beta\gamma} U^\alpha U^\delta+\eta^{\alpha\delta} U^\gamma U^\beta+\eta^{\delta\gamma} U^\beta U^\alpha+\eta^{\delta\beta} U^\alpha U^\gamma\right) \nonumber \\
&&+\frac{2 K_2(\chi)}{\chi^2}(\eta^{\alpha\beta}\eta^{\gamma\delta}+\eta^{\alpha\gamma}\eta^{\beta\delta}+\eta^{\alpha\delta}\eta^{\beta\gamma}),
\end{eqnarray}
Similarly, ${Z^\star}^{\alpha\beta\gamma\delta...}=\int_{-\infty}^\infty P^\alpha P^\beta P^\gamma P^\delta ... \exp(-\chi P^\alpha U_\alpha) \frac{dP}{P^0}$ is obtained by the successive differentiation of $Z^\star$ using $-\chi U_\alpha$ as
\begin{eqnarray}
&&Z^\star=2 K_0\left(Q^\star \chi \right),\\
&&{Z^\star}^\alpha=2 Q^\star K_1(Q^\star \chi)U^\alpha,\\
&&{Z^\star}^{\alpha\beta}=2 {Q^\star}^2 K_2(Q^\star \chi)U^\alpha U^\beta-2 Q^\star \eta^{\alpha\beta} \frac{K_1(Q^\star \chi)}{\chi},
\end{eqnarray}
\end{appendix}
\end{document}